\documentclass{elsart}

 \usepackage{epsfig}

\usepackage{amssymb}

\begin{document}

\begin{frontmatter}

\title{Thermodynamic and transport properties of RAgGe
(R=Tb-Lu) single crystals}

\author[a1,a2]{E. Morosan,}
\author[a1]{S. L. Bud'ko,}
\author[a1,a2]{P. C. Canfield,}
\author[a3]{M. S. Torikachvili,}
\author[a4]{A. H. Lacerda}

\address[a1]{Ames Laboratory U.S. D.O.E., Iowa State University, Ames, IA 50011}
\address[a2]{Department of Physics and Astronomy, Iowa State University, Ames, IA 50011}
\address[a3]{Department of Physics, San Diego State University, San Diego, CA 92182}
\address[a4]{National High Magnetic Field Laboratory, Los Alamos Facility, Los Alamos National Laboratory, Los Alamos, NM 87545}

\begin{abstract}
Single crystals of the title compounds were grown out of an
AgGe-rich ternary solution. Powder x-ray diffraction data
confirmed the hexagonal AlNiZr-type structure ($P\bar{6}2m$  space
group), an ordered variant of the Fe$_2$P structure type.
Antiferromagnetic ordering can be inferred from magnetization,
resistance and specific heat measurements, with values of T$_N$
between 28.5 K for TbAgGe, and 1.0 K for YbAgGe, which scale
roughly with the de Gennes factor. Anisotropic $M(H)$ measurements
indicate one or more metamagnetic transitions when the external
field is applied along the $c$-axis (for R=Tb) or perpendicular to
it (R = Ho, Er, Tm), or even in both orientations as in the case
of DyAgGe. Furthermore, the extreme anisotropy of the
magnetization in TmAgGe, where magnetic moments lie in the
$ab$-plane, provides the possibility of studying the angular
dependence of metamagnetism in hexagonal compounds with the rare
earth in orthorhombic point symmetry.

YbAgGe has distinct properties from the rest of the series: an
enhanced electronic specific heat coefficient ($\gamma
\approx$(154.2$\pm$2.5)mJ/mol*K$^2$), and apparently small moment
magnetic ordering below 1.0 K. This compound appears to be close
to a quantum critical point.
\end{abstract}

\end{frontmatter}

\section{Introduction}
Ternary intermetallic compounds R-T-M, with R=rare earth metals,
T=transition metals, M=metals of the $p$ block, have raised a lot
of interest in the past years, given their structural complexity
and their greatly varying physical properties.  Studies of the
anisotropic properties of such materials, with the R in tetragonal
point symmetry, revealed anisotropy (in some cases extreme) in
members of the RAgSb$_2$ \cite{1} and RNi$_2$Ge$_2$ \cite{2}
series, as well as in the well known quaternary RNi$_2$B$_2$C
compounds, for R=Tb-Er \cite{3,4,5,6}. The strong crystalline
electric field (CEF) anisotropy confines the moments either along
the $c$-axis of the tetragonal unit cell (i.e. in TbNi$_2$Ge$_2$,
ErAgSb$_2$, TmAgSb$_2$), or to the basal plane $ab$ (in
ErNi$_2$Ge$_2$, DyAgSb$_2$ and the aforementioned RNi$_2$B$_2$C
compounds), and metamagnetic transitions occur in the majority of
these materials. TbNi$_2$Ge$_2$ was chosen for an extensive study
of angular dependent metamagnetism in an axial (Ising-like)
compound \cite{2}, showing fairly simple angular dependencies of
locally saturated magnetizations ($M_{sat}(\theta)$) and critical
fields ($H_c(\theta)$). The complexity of the $M_{sat}(\theta)$)
and $H_c(\theta)$ phase diagrams drastically increases for systems
where the magnetic moments are confined to the $ab$-plane, as
shown by the detailed analysis of $M(H,\theta)$ data done for
HoNi$_2$B$_2$C \cite{3} and DyAgSb$_2$ \cite{7}. Furthermore, a
theoretical model has been developed by Kalatsky and Pokrovsky
\cite{8}, to explain the experimentally observed behavior of
metamagnetism in HoNi$_2$B$_2$C. An appropriate variant of this
four-position clock model subsequently agreed with the data on
DyAgSb$_2$ \cite{7}.

In addition to metamagnetism, hybridization of the $4f$ moments
occurs in some compounds, making these tetragonal series even more
interesting: YbNi$_2$B$_2$C \cite{9}, CeNi$_2$Ge$_2$ \cite{2,10},
YbNi$_2$Ge$_2$ \cite{2} are reported to have significant
hybridization between the 4f and the conduction electrons.

Having achieved a basic understanding of the physical properties
of these tetragonal compounds, in which the R$^{3+}$ ions are
positioned in crystallographically unique tetragonal point
symmetry sites, we anticipate that materials with different
crystal structure would be of further interest. Crystals with
hexagonal unit cells  preserve the axial versus basal plane
anisotropy while allowing for three interesting R point
symmetries: orthorhombic, trigonal or hexagonal.  A representative
of the first class of materials (i.e. orthorhombic point symmetry)
is the RAgGe series, that crystallizes in the AlNiZr structure, an
ordered variant of the Fe$_2$P structure.

Detailed structural studies of RAgGe for the heavy rare earth
members were reported in refs. \cite{11} and \cite{12}.  The
R$_3$Ag$_3$Ge$_3$ structure can be viewed as alternating R$_3$Ge
and Ag$_3$Ge$_2$ layers stacked along the $c$ axis, shown as a
$c$-axis projection in Fig. \ref{f1}. In this structure there are
three rare earth atoms in the hexagonal unit cell, sitting at $3g$
sites with orthorhombic symmetry. Using single crystal samples we
have done a detailed analysis of temperature and field dependent
anisotropic magnetization and transport measurements, and
established the trends caused by the change of the rare earth in
the physical properties. This has allowed us to more fully
characterize the series than in the previous studies done on
polycrystalline samples \cite{12,13}, and establish an
experimental baseline for more detailed studies of specific
members of this series.

In this paper we will present a brief review of experimental
methods, followed by detailed experimental results of
magnetization, magnetotransport and specific heat measurements for
each compound.  More attention will be paid to (i) YbAgGe, for
which additional measurements down to 0.4 K were performed, and
(ii) TmAgGe, for which an initial study of the in-plane anisotropy
will be presented. The experimental section is followed by a brief
discussion, where we will emphasize the observed trends in the
magnetic properties within the series, as well as a few
outstanding questions we are presently trying to address.

\section{Experimental methods}

\begin{figure}
\begin{center}
\includegraphics[angle=0,width=120mm]{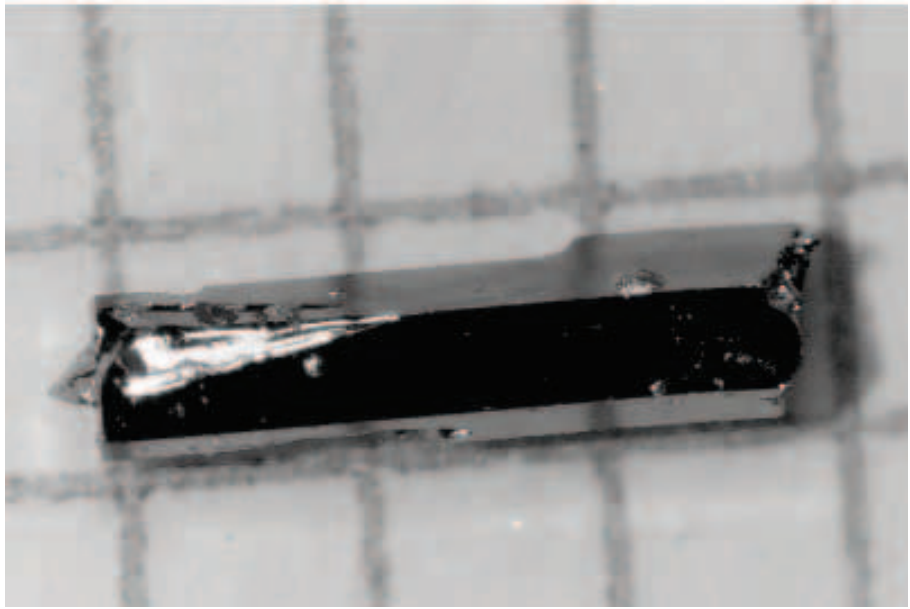}
\end{center}
\caption{Single crystal of YbAgGe, with approximate dimensions
$0.4 \times 0.4 \times 3.0$ mm. Hexagonal rod geometry evident
(three of the six possible facets are visible here); a few AgGe
flux droplets can be seen on the surface of the
crystal.}\label{f2}
\end{figure}

\begin{figure}
\begin{center}
\includegraphics[angle=0,width=120mm]{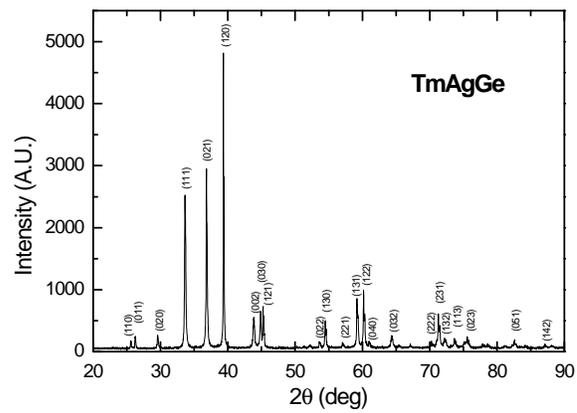}
\end{center}
\caption{Powder X-ray diffraction pattern for TmAgGe. Peaks are
indexed to a hexagonal structure, with $a=7.0565 \AA$ and
$c=4.1454 \AA$.}\label{f3}
\end{figure}

Single crystals of RAgGe for R = Tb-Lu were grown from high
temperature ternary solutions, rich in Ag and Ge \cite{14,15,16}.
Typical concentrations that resulted in good, well formed single
crystals were R$_x$(Ag$_{0.75}$Ge$_{0.25}$)$_{1-x}$, with
$x$=0.06-0.14. This Ag-Ge rich self-flux was preferred not only
because of the low-temperature, binary eutectic
($\sim$650$^\circ$C) around a Ag concentration of 75\%, but also
because it introduces no additional elements to the melt. The
constituent elements were placed in an alumina crucible and sealed
in a quartz ampoule under a partial pressure of Ar. After
initially heating the ampoule to above $\sim$1100$^\circ$C, it was
slowly cooled to 850-825$^\circ$C.  After the slow cooling the
excess solution was decanted and thin hexagonal rods, with the
$c$-axis along the axis of the rod were obtained.  Two noteworthy
modifications to this growth procedure were YbAgGe which was
cooled to 750$^\circ$C before decanting and TbAgGe for which an
initial melt stoichiometry of
Tb$_x$(Ag$_{0.85}$Ge$_{0.15}$)$_{1-x}$ was necessary for better
quality crystals.  Fig. \ref{f2} shows a picture of an YbAgGe
single crystal. Powder X-ray diffraction measurements were taken
at room temperature with Cu K$_\alpha$ radiation ($\lambda$ =
1.5406 $\AA$) in order to confirm the crystal structure and to
check for impurity phases. A typical pattern is shown in Fig.
\ref{f3} for TmAgGe. There are no detectable second phase peaks
and all of the detected ones can be indexed using a hexagonal
structure with a=7.05(0.01) $\AA$ and c=4.14(0.01) $\AA$. In Fig.
\ref{f4}, the volume and the dimensions of the unit cell across
the series are shown as a function of the R$^{3+}$ ionic radii
\cite{17}, illustrating the expected lanthanide contraction.
YbAgGe falls close to the monotonic decrease of the unit cell
parameters, consistent with  the Yb ion being essentially
trivalent at room temperature in this compound.

Attempts were made to grow even lighter rare earth members of this
series (including YAgGe), both around and off the eutectic point
of the Ag-Ge binary phase diagram. For R=Gd, we failed to grow the
right phase when introducing rare earth amounts varying from 6\%
to as much as 37.5\% into the Ge$_{0.25}$Ag$_{0.75}$ melt; instead
an initial composition of
Gd$_{0.06}$(Ge$_{0.25}$Ag$_{0.75}$)$_{0.94}$ yielded orthorhombic
crystals of Gd$_3$Ag$_4$Ge$_4$, space group $Immm$ and lattice
parameters, as given by single crystal structure determination,
$a$=4.34 $\AA$, $b$=7.07 $\AA$ and $c$=14.49 $\AA$.  All the other
compositions or R that we tried resulted in either no
crystallization or crystals of as of yet unidentified phases.

The magnetization measurements on the RAgGe compounds that we are
reporting here were performed in a Quantum Design MPMS SQUID
magnetometer ($T$=1.8–350K, $H_{max}$= 55 or 70kG). We measured
anisotropic magnetization for all compounds, having the external
field $H\|c$ or $H\|ab$ (in a random {\it in-plane} orientation,
unless otherwise specified), and the corresponding
susceptibilities were then calculated as $\chi=M/H$; the
polycrystalline average susceptibility was evaluated as
$\chi_{ave} = (\chi_c+2*\chi_{ab})/3$. We will infer the values of
the transition temperatures (N\'{e}el temperature or spin
reorientation temperatures) as determined from $d(\chi*T)/dT$
\cite{18} plots as well as from plots of $C_p(T)$ and $d\rho/dT$
\cite{19}. YbAgGe is an exception and the criterion for
determining its transition temperature will be discussed in the Yb
section.

Additional measurements of the high field DC magnetization of
TbAgGe were carried out with a vibrating sample magnetometer in a
180kG superconducting magnet (National High Magnetic Field
Laboratory - Los Alamos Facility). For DyAgGe, HoAgGe and YbAgGe,
$M(H)$ measurements up to $H$=140kG were performed in a Quantum
Design PPMS system using the extraction magnetization option. Most
of the samples were manually aligned to measure magnetization
along the desired axis, but for $M(H,\theta)$ measurements in
TmAgGe, the angular position of the sample was controlled by a
modified MPMS sample rotor.

The electrical resistance in zero and applied field was measured
with a Linear Research LR-700 AC resistance bridge ($f$=16Hz,
$I$=1-3mA) in the magnetic field –temperature environment of the
QD MPMS system, using a standard four-probe technique. The current
was flowing along the $c$-axis of the crystals. Due to the
limitations imposed by the thin-rod shape of the samples, for
axial ($H \| c$) applied magnetic field, {\it longitudinal} ($H \|
I$) magnetoresistance was measured, whereas for magnetic field
applied in the basal plane ($H \| ab$), {\it transverse} ($H \perp
I$) magnetoresistance was studied. A $^3$He cooling system in the
QD PPMS was used for measurements of the resistance of YbAgGe down
to low temperatures ($T$=0.4K).  Given the geometry of the samples
(partial or full hexagonal cross sections) and the orientation of
the sample probe with respect to the applied field, the field
direction could in some cases be inferred to be $H \| [120]$ (when
using an orthogonal set of coordinates with $a'=a$ and $b' =
a*\frac{\sqrt{3}}{2}$ while a is the hexagonal unit cell
parameter). This will be discussed in further detail in the
appropriate sections.

Heat capacity measurements were made in a Quantum Design PPMS
system; in the cases of YbAgGe and LuAgGe, measurements down to
$T$=0.4K were done using the $^3$He cooling option. The sample
holder and grease background, measured separately for each sample,
was later subtracted from the sample response. In order to
estimate the magnetic contribution to the specific heat for the
moment-bearing rare earth compounds, the specific heat of LuAgGe
was subtracted. The magnetic part of the specific heat was used
for the numerical integration of $C_m(T)/T$ vs. $T$, which allowed
us to estimate the temperature dependence of the magnetic entropy
for each compound.  The low temperature data (below actual
measurements) for $C_p(T)$ were estimated by non-linear fitting to
(0,0) under the assumption that the small error thus introduced is
further minimized by the subtraction of the Lu analogue.

\section{Results}
In presenting the data, we are characterizing each compound by
magnetization, electrical resistivity and specific heat
measurements. We will start with LuAgGe as the non-magnetic member
of the series, and then progress from R=Tb through Yb. For each
material and measurement that we present, we will try to emphasize
properties that we find to persist in the series, as well as
specific characteristics of each compound.

\subsection{LuAgGe}

\begin{figure}
\begin{center}
\includegraphics[angle=0,width=120mm]{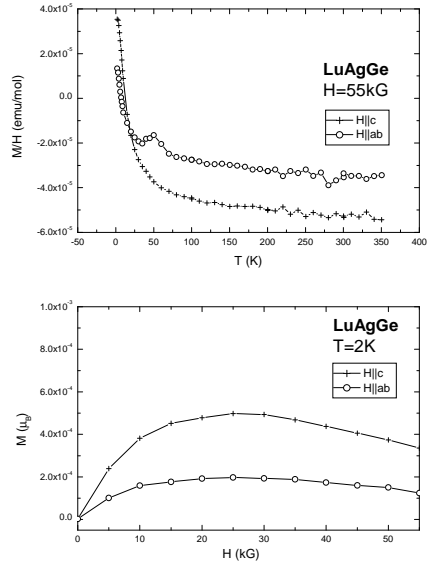}
\end{center}
\caption{(a) Anisotropic temperature-dependent susceptibility and
(b) anisotropic magnetization isotherms at $T$=2K of LuAgGe.
}\label{f5}
\end{figure}

LuAgGe has electronic and magnetic properties consistent with a
weakly diamagnetic intermetallic compound with no magnetic order.
The magnetization as a function of temperature (Fig. \ref{f5}a) is
almost constant, with a very small, average, high temperature
values around -2.3$\times 10^{-5}$ emu/mol ($H \| {ab}$), and
-3$\times 10^{-5}$ emu/mol ($H \| c$) respectively. At low
temperatures, an upturn in the susceptibility data occurs, which
could be a consequence of some magnetic impurities being present
in the original materials ({\it e.g.} the tail indicates a
magnetic impurity contamination equivalent to 0.2\% Gd). Field
dependent magnetization curves for the two orientations of the
field are shown in Fig. \ref{f5}b; as expected for a non-magnetic
compound with some magnetic impurities, the magnetization is a
superposition of Brillouin saturation of the impurities with
applied field, and a weak diamagnetic signal.

\begin{figure}
\begin{center}
\includegraphics[angle=0,width=120mm]{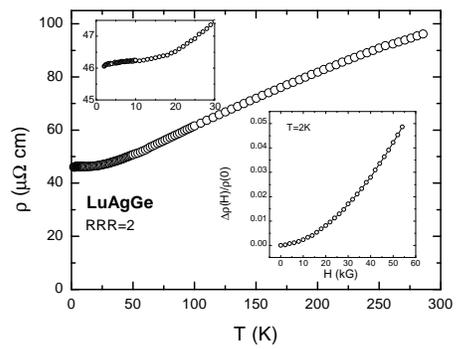}
\end{center}
\caption{Zero-field resistivity (upper left insert - low
temperature part) and transverse magnetoresistance (lower right
inset) of LuAgGe.}\label{f6}
\end{figure}

The temperature dependent resistivity of LuAgGe (Fig. \ref{f6})
demonstrates the metallic character of this compound. Below
$\sim$50K, the impurity scattering becomes dominant, with a
relatively large residual resistivity $\rho(1.8K)\approx
45\mu\Omega cm$, resulting in a relatively poor RRR of about 2,
far smaller than the RRR values found in the rest of the series.
The upper inset to Fig. \ref{f6} shows that there is no clear
resistance minimum found in LuAgGe at low temperature. This is
important to note given that for the R=Tb-Tm members of the RAgGe
series do manifest a minimum in the temperature dependent
resistivity for temperatures well above their respective N\'{e}el
temperatures. The lower inset to Fig. \ref{f6} presents the
transverse magnetoresistance, which varies approximately as $H^2$,
as expected for normal metals.

\begin{figure}
\begin{center}
\includegraphics[angle=0,width=120mm]{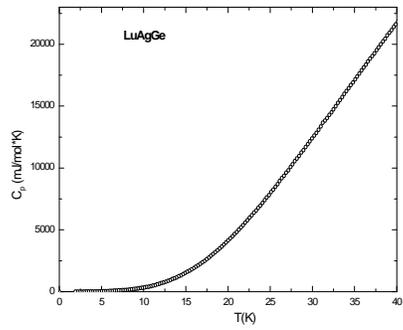}
\end{center}
\caption{Heat capacity of LuAgGe.}\label{f7}
\end{figure}

\begin{figure}
\begin{center}
\includegraphics[angle=0,width=120mm]{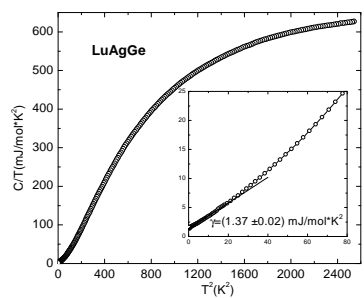}
\end{center}
\caption{$C_p/T$ versus $T^2$ for LuAgGe; insert: low temperature
part, line in the inset is a linear fit at low
temperatures.}\label{f8}
\end{figure}

Finally, we have measured the specific heat $C_p$ as a function of
temperature (Fig. \ref{f7}). These data will be used as an
estimate of the non-magnetic contribution to the specific heat in
all RAgGe, R=Tb-Yb. As calculated from the linear fitting of the
low-temperature $C_p/T(T^2)$ data (Fig. \ref{f8}), the electronic
specific heat coefficient for the non-magnetic LuAgGe compound is:
 $\gamma$=(1.37$\pm$0.02) mJ/mol*K$^2$ whereas  $\Theta_D
\approx$ 300 K.

\subsection{TbAgGe}

\begin{figure}
\begin{center}
\includegraphics[angle=0,width=120mm]{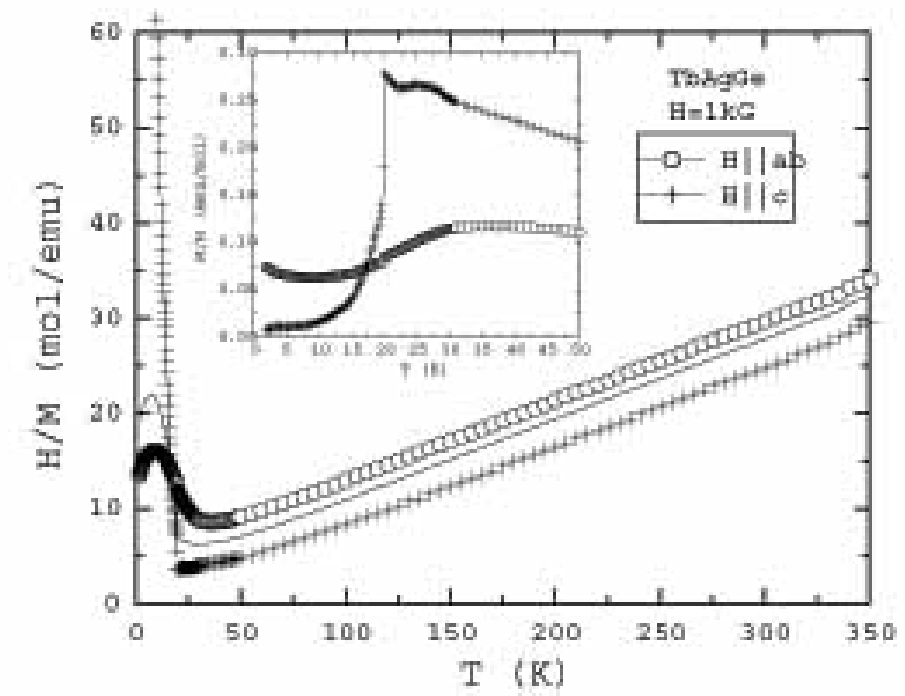}
\end{center}
\caption{Anisotropic inverse susceptibilities of TbAgGe and
calculated average (line); inset: low-temperature anisotropic
susceptibilities.}\label{f9}
\end{figure}

The anisotropic, inverse susceptibility as a function of
temperature for TbAgGe is shown in Fig. \ref{f9}, together with
the calculated polycrystalline average.  The inset presents the
low temperature region of the susceptibility for the applied field
$H$=1kG parallel and perpendicular to the $c$-axis.  The inverse
susceptibility above $\sim$50K is consistent with Curie-Weiss
behavior of the magnetization: $\chi(T)=C/(T+\Theta_P)$, where
$\Theta_P$ is the paramagnetic Weiss temperature. The values of
$\Theta_P$ for the two orientations of the field, as well as for
the polycrystalline average, are listed in Table \ref{t1} in the
discussion section.  A linear fitting of the average inverse
susceptibility in the paramagnetic state gives a value of the
effective moment $\mu_{eff} = 9.7 \mu_B$, which is close to the
theoretical value for Tb$^{3+}$ of 9.72$\mu_B$.

\begin{figure}
\begin{center}
\includegraphics[angle=0,width=120mm]{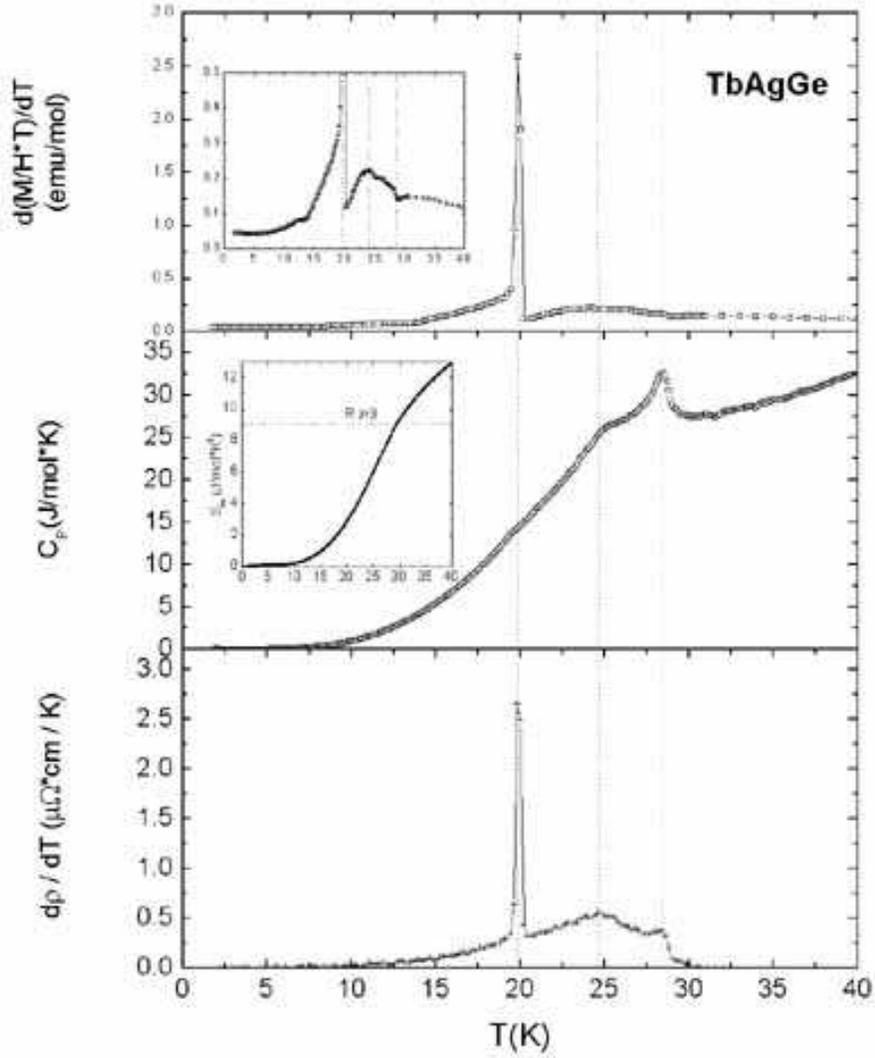}
\end{center}
\caption{(a)Low-temperature $d(\chi T)/dT$ for TbAgGe (inset:
enlarged to show lower peaks' position in $d(\chi T)/dT)$; (b)
specific heat $C_p(T)$ with the magnetic entropy $S_m$ in the
inset; (c) low-temperature $d\rho/dT$; dotted lines mark the peak
positions as determined from (a).}\label{f10}
\end{figure}

The low temperature susceptibility for $H$=1kG (shown as an inset
in Fig. \ref{f9}) indicates antiferromagnetic ordering below the
N\'{e}el temperature $T_N$=28.4K. This transition temperature, as
well as two others are seen more clearly for $H \| c$ and can also
be identified in the $d(\chi T)/dT$ (around 28.5K, 24.7K and 19.8K
respectively) and $d\rho/dT$ (at 28.4K, 24.6K and 19.8K) (Fig.
\ref{f10}a and c); in $C_p(T)$ (Fig. \ref{f10}b) only the higher
two transitions are visible, around 28.3K and 24.7K respectively.
Another possible change in slope in $M(T)/H$  occurs around 18K,
but it is obscured in all other measurements, and is therefore
unclear if it can be associated with another magnetic transition.
However, in the $H_c(T)$ phase diagram for $H \| c$ (shown below),
we can follow a phase boundary indicative of such a transition.

Specific heat data was also used to estimate the magnetic entropy
$S_m(T)$, shown in Fig. \ref {f10}b, inset.  From these data it
can be inferred that the ordered state in TbAgGe is emerging out
of triplet ground state, or there are at least a combination of a
singlet and a doublet, or three singlet states closely spaced.

\begin{figure}
\begin{center}
\includegraphics[angle=0,width=120mm]{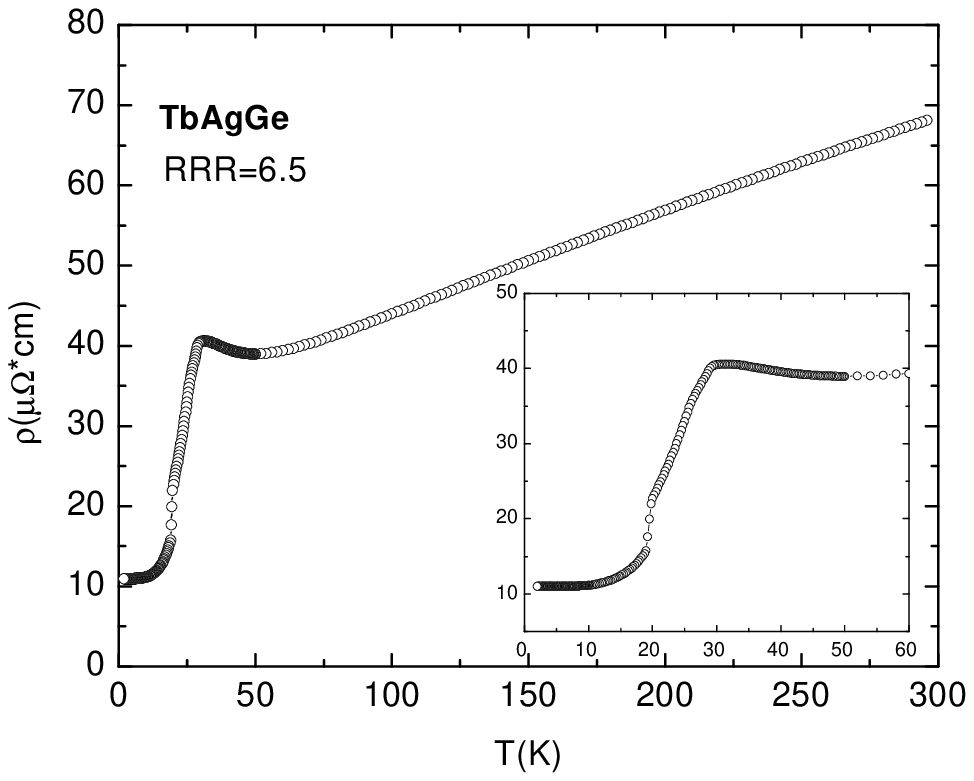}
\end{center}
\caption{Zero-filed resistivity of TbAgGe (inset: enlarged
low-temperature part).}\label{f11}
\end{figure}

At high temperatures, the resistivity measurements (Fig.
\ref{f11}) indicate the metallic character of TbAgGe as $\rho$
monotonically decreases with decreasing $T$. The residual
resistivity ratio RRR, calculated as $\rho(300K)/\rho(2K)$ and
equal to 6.5, indicates fair crystal quality. The low temperature
region of the resistivity measurements features slope changes
associated with the magnetic transitions discussed above. On the
other hand, there is a clear minimum of the $\rho(T)$ curve around
$T \approx$ 50K. This minimum in resistivity occurs far above the
N\'{e}el temperature (around 2*$T_N$), which rules out the
possibility of a superzone gap causing this feature, and, as will
be seen below, occurs for all other local moment-bearing members
of the series. This feature could be explained by magnetic
fluctuations or some other, as of yet to be identified, mechanism.

Based on all the measurements performed on single crystals of
TbAgGe, we can identify at least three transitions, at
T=(28.4$\pm$0.1)K, (24.65$\pm$0.05)K and (19.8$\pm$0.1)K; the
lowest two temperatures are very close to those reported in
\cite{13} for the polycrystalline samples (from magnetization and
neutron diffraction measurements) as the only transitions. We are
thus led to believe that the highest (antiferromagnetic)
transition temperature, as well as any other possible ordering
temperatures, was not detected by the measurements made on
polycrystalline samples.

\begin{figure}
\begin{center}
\includegraphics[angle=0,width=120mm]{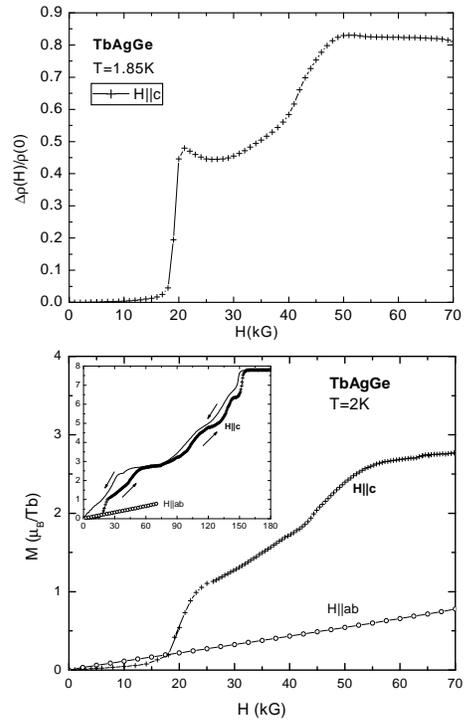}
\end{center}
\caption{(a) Transverse magnetoresistance for TbAgGe; (b)
anisotropic magnetization curves for $H$ up to 70kG (inset:
$M(H,T$=2K$)$ for increasing and decreasing field $H\|c$, up to
$H$=180kG).}\label{f12}
\end{figure}

Both resistivity (Fig. \ref{f12}a) and magnetization measurements
(Fig. \ref {f12}b) as function of applied field provide evidence
for a series of metamagnetic transitions in TbAgGe. Initial field
dependent magnetization and resistivity measurements up to
$H$=70kG reveal at least two metamagnetic transitions for critical
fields $H_c \approx$ 20kG, and 48kG respectively, when field is
applied along the $c$-axis.  Given that the system seemed to be
far below saturation ($M(70kG)=2.77\mu_B/$Tb $\ll$ 9$\mu_B/$Tb),
further magnetization measurements were performed for fields up to
180kG and several other metamagnetic transitions were observed.
(It should be noted that there is clear indication of hysteresis
as manifested by the difference in $M(H)$ for increasing and
decreasing field measurements (inset, Fig. \ref{f12}b)). The value
of the magnetization at 180kG is $M=7.78 \mu_B/$Tb, still below
the calculated value for $\mu_{sat}($Tb$^{3+})=9.0\mu_B$. This
somewhat suppressed value of $M$ could be consistent with more
metamagnetic states beyond 180kG, or with moments alignment at
some angle $\phi \not= 0$ with respect to the $c$-axis, or may
simply be a caliper of the uncertainty in the absolute value of
$M$ in this high field measurement.

\begin{figure}
\begin{center}
\includegraphics[angle=0,width=120mm]{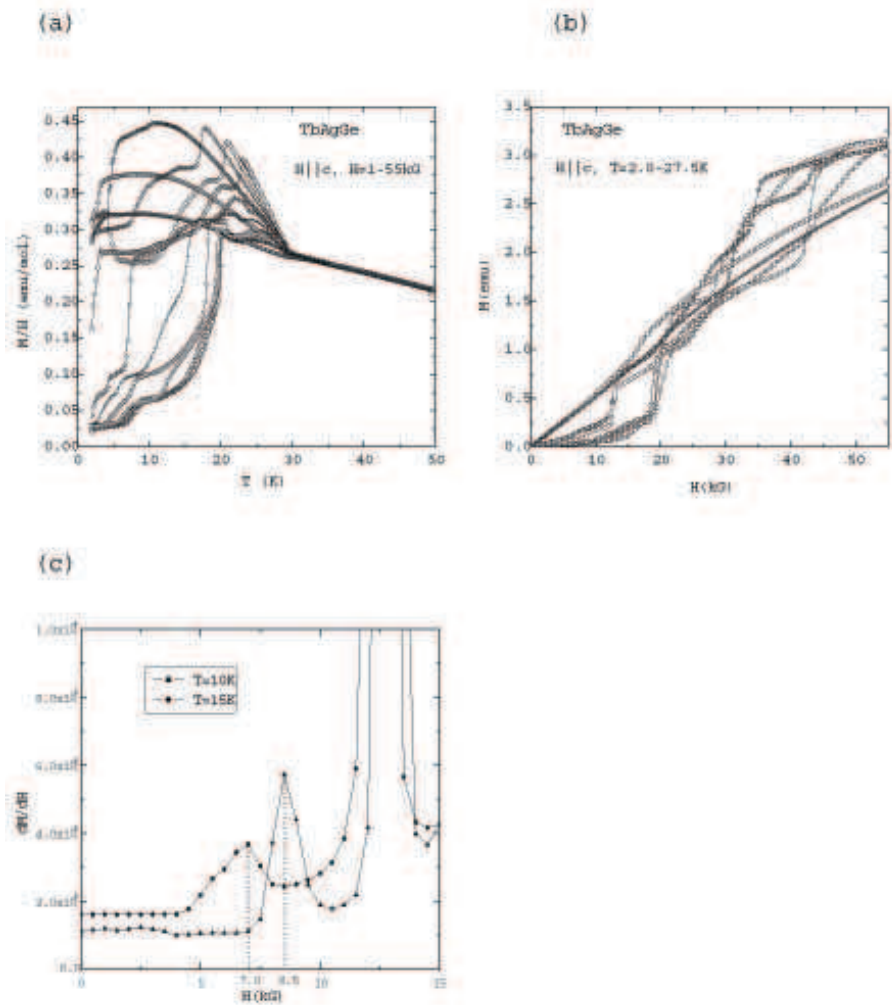}
\end{center}
\caption{$\chi(T)$ curves for various (1, 2.5, 5, 7.5, 10, 12.5,
15, 20, 25, 30, 35, 45, 55 kG) fields (a) and several $M(H)$ ($T$
= 2, 3, 4, 6, 10, 20, 25, 27, 50 K) isotherms (b); (c) - enlarged
derivatives of $M(H)$ (for $T$=10K and 15K) as an example of how
the lower line in the phase diagram was obtained.}\label{f13}
\end{figure}

\begin{figure}
\begin{center}
\includegraphics[angle=0,width=120mm]{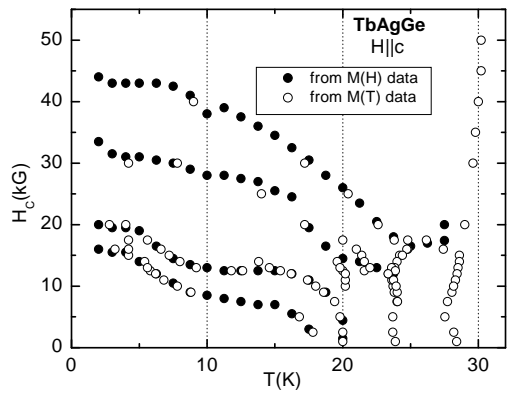}
\end{center}
\caption{$H_c-T$ phase diagram as determined from the
magnetization curves in Fig. 13.}\label{f14}
\end{figure}

In order to better determine the number and extent of metamagnetic
phases that exist for TbAgGe when $H \| c$, more measurements were
done (Fig. \ref{f13}a,b), which allowed us to plot a tentative
$H_c(T)$ phase diagram shown in Fig. \ref{f14}. Local maxima of
$d(\chi T)/dT$ or $dM/dH$ were used to estimate the $H_c$ and
$T_c$ values, as illustrated in Fig. \ref{f13}c. Numerous phases
can be observed. Whereas figure \ref{f14} clearly shows the three,
$H = 0$, transition temperatures discussed above, it also shows a
lower field and temperature phase line existing for finite applied
fields.  Figure \ref{f13}c illustrates that this lowest phase line
is clearly detected in our $M(T,H)$ data.  A remaining question
associated with this phase diagram is whether this lowest phase
line intersects the $H = 0$ axis at any finite temperature or
flattens out at very low fields such that it intersects $H=0$ at
$T \approx $20 K, where another transition already exists.

\subsection{DyAgGe}
\begin{figure}
\begin{center}
\includegraphics[angle=0,width=120mm]{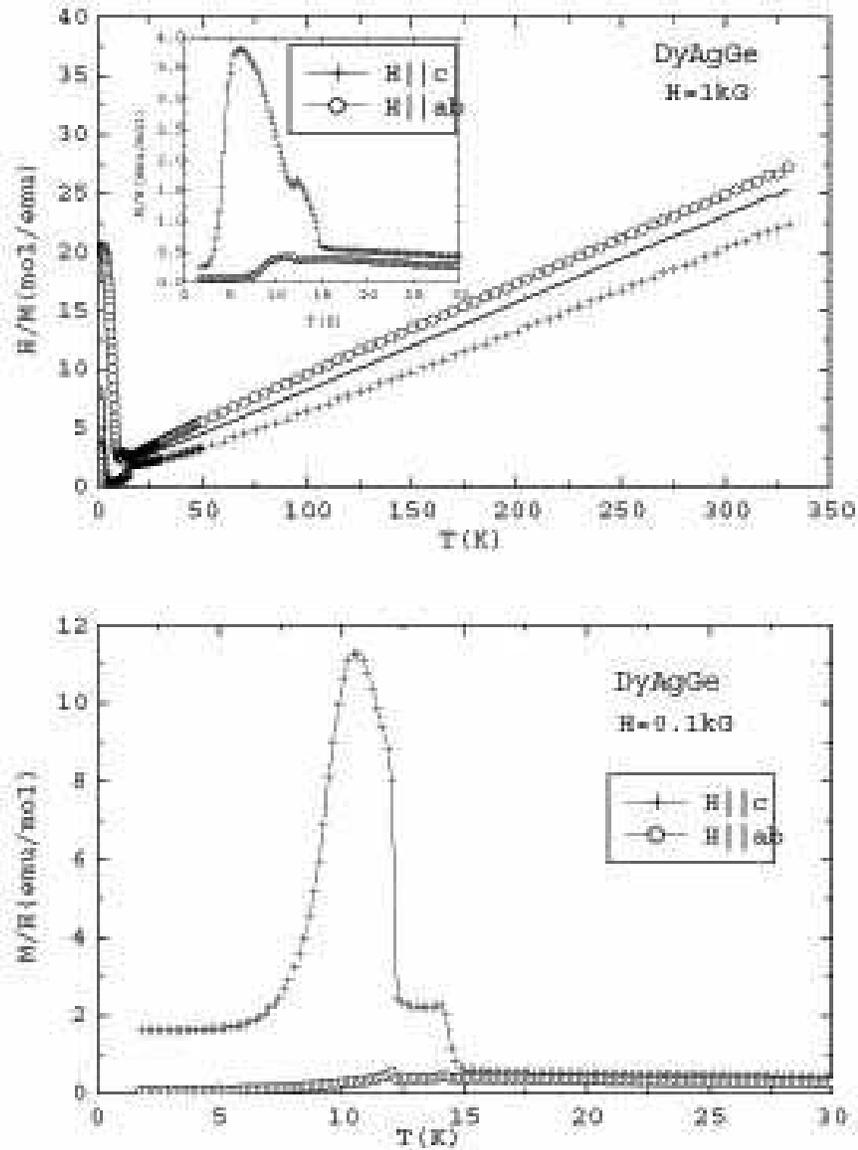}
\end{center}
\caption{(a) Anisotropic inverse susceptibilities of DyAgGe and
calculated average (line) at H=1kG; inset: low-temperature
anisotropic susceptibilities; (b) low-temperature anisotropic
susceptibilities for H=0.1kG.}\label{f15}
\end{figure}

\begin{figure}
\begin{center}
\includegraphics[angle=0,width=120mm]{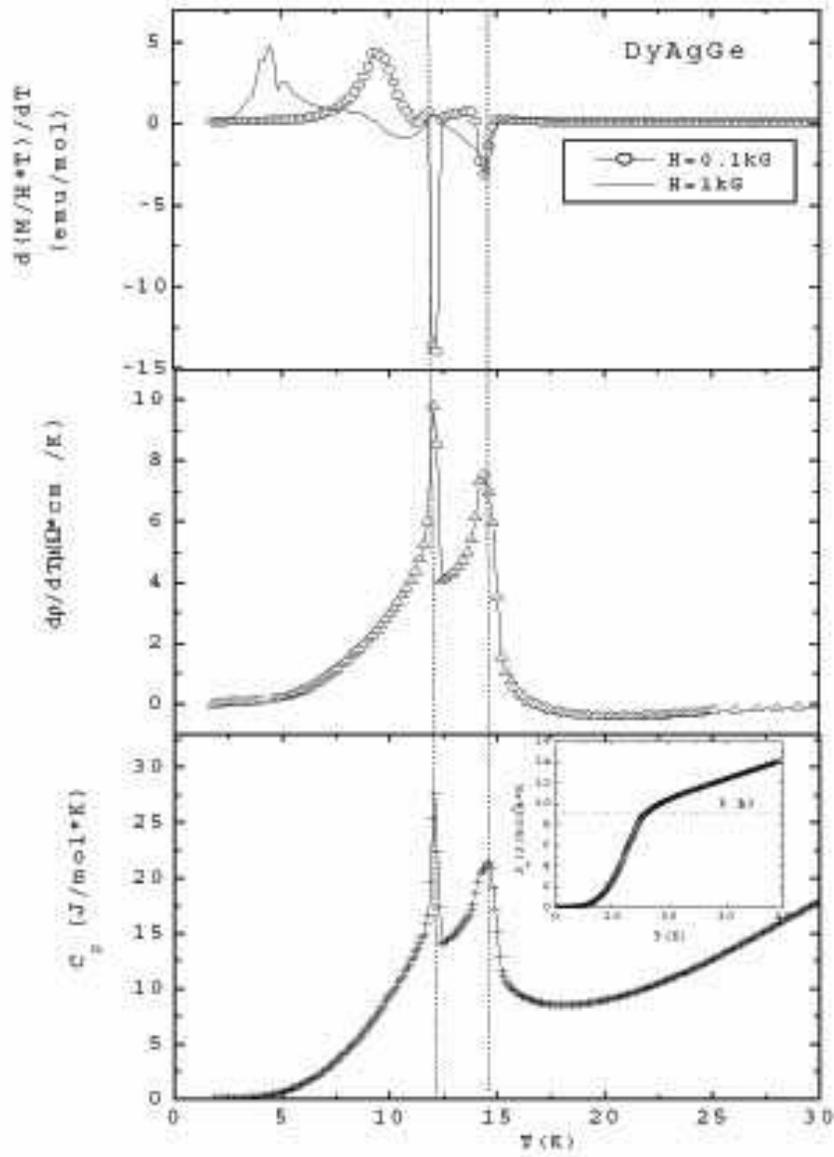}
\end{center}
\caption{(a) Low-temperature $d(\chi T)/dT$ for TbAgGe for
$H$=0.1kG and 1kG; (b) low-temperature $d\rho/dT$; (b) specific
heat $C_p(T)$ with the magnetic entropy $S_m$ in the inset; dotted
lines mark the peak positions as determined from lowest field data
in (a). }\label{f16}
\end{figure}

\begin{figure}
\begin{center}
\includegraphics[angle=0,width=120mm]{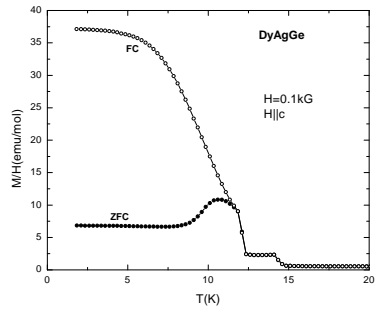}
\end{center}
\caption{ZFC-FC magnetization of DyAgGe, $H$=0.1kG
$(H\|c)$.}\label{f17}
\end{figure}

Although DyAgGe in the paramagnetic state is more isotropic than
TbAgGe, with slightly larger susceptibility for $H \| c$ than for
$H \| ab$ (Fig. \ref{f15}), it has an extremely anisotropic
ordered state (as seen in Fig. \ref{f15}a inset, for $H$=1kG);
this low temperature anisotropy is further enhanced for lower
(0.1kG) applied fields (Fig. \ref{f15}b). Two transition
temperatures are detected by $d(\chi T)/dT$ (at 14.4 and 12.0 K),
$d\rho/dT$ (at 14.4K, and 12.0K) and $C_p(T)$ (at 14.6K and 12.1K)
(Fig. \ref{f16}). These are in agreement with the temperatures
determined by earlier measurements on polycrystalline samples
\cite{12,13}.  There is one more, broader, lower temperature peak
in Fig. \ref{f16}a, not visible in any other measurement. But
comparison of $d(\chi T)/dT$ plots for two different applied
fields (Fig. \ref{f16}a, $H$=0.1kG and $H$=1kG) seems to indicate
that the low temperature peak is moving down in $T$ rapidly as the
applied field increases, whereas the other two are unaffected by
the change in $H$.  In order to examine the low temperature state
of DyAgGe more carefully 0.1 kG zero-field-cooled-warming (ZFC)
and field-cooled warming (FC) data sets were taken and are plotted
in Fig. \ref{f17}. These data are consistent with an ordered state
below 12.0 K that has a net ferromagnetic component along the
$c$-axis. The broadness and the field sensitivity of the lowest
temperature peak in Fig. \ref{f16}a, as well as the difference
between the ZFC and FC data sets shown in Fig. \ref{f17} are
consistent with the rotation of domains in small applied fields.
It should also be noted that the magnetization associated with the
lowest temperature point of the FC curve shown in Fig. \ref{f17}
corresponds to $M \sim 0.66 \mu_B/$Dy, the value associated with
the low field plateau of the $M(H)$ plot shown in Fig. \ref{f18}.
This further supports the idea that below $T \sim$ 12 K the
magnetically ordered state of DyAgGe has a net ferromagnetic
component along the $c$-axis.

The linear behavior of the inverse susceptibility above $\sim$ 50K
(Fig. \ref{f15}a) is indicative of Curie-Weiss like
susceptibility, the effective magnetic moment determined from the
linear region being $\mu_{eff} = 10.3 \mu_B$ (in good agreement
with the theoretical value for Dy$^{3+}$, which is 10.6$\mu_B$).
Paramagnetic Weiss temperatures $\Theta_P$ are listed in Table
\ref{t1}.

\begin{figure}
\begin{center}
\includegraphics[angle=0,width=120mm]{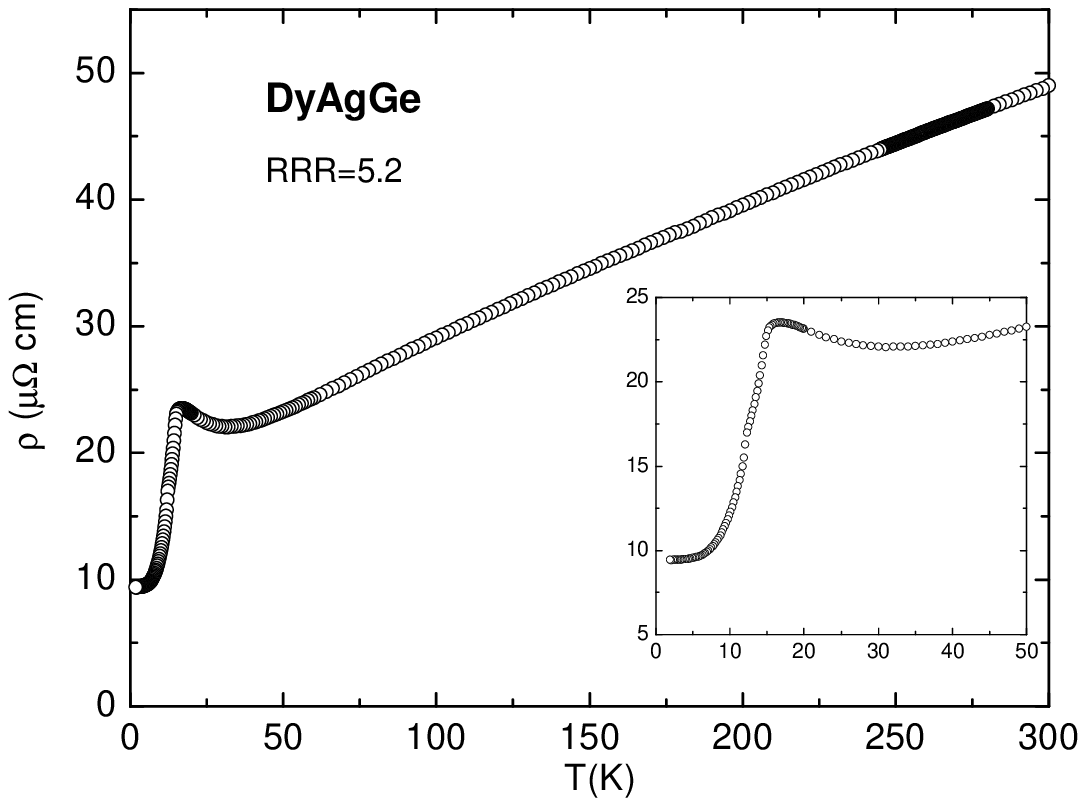}
\end{center}
\caption{Zero-filed resistivity of DyAgGe (inset: enlarged
low-temperature part).}\label{f19}
\end{figure}

The resistance of DyAgGe increases linearly with temperature above
50K (Fig. \ref{f19}). At low temperatures, the resistivity has a
local minimum around 30K, followed by a sharp drop around the
transition temperature; this indicates loss of spin disorder
scattering as the system enters the ordered state. The residual
resistivity ratio RRR is approximately 5.2.

Field dependent magnetization measurements indicate a complex
metamagnetism in DyAgGe, as shown in Fig. \ref{f18}: transitions
can be seen in all three orientations of the applied field ($H \|
[001]$, $H \| [120]$ and $H \| [010]$), which probably means that,
in the saturated state, the magnetic moments are inclined at some
angle $0^\circ < \varphi < 90^\circ$ with respect to the $c$-axis.
This could justify the low magnetization values (below the
theoretical value $\mu_{sat}($Dy$^{3+})=10\mu_B$) in all three
directions even at the highest applied field, as we explain in
more detail in the discussion section.

\begin{figure}
\begin{center}
\includegraphics[angle=0,width=120mm]{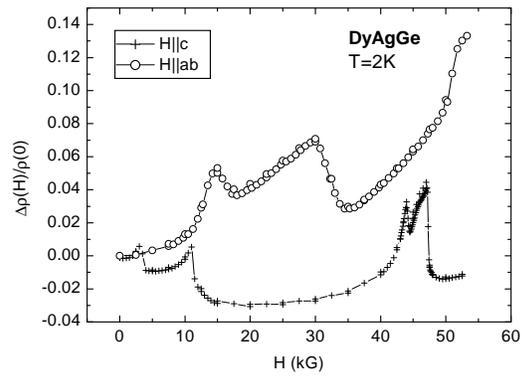}
\end{center}
\caption{Anisotropic magnetoresistance at T=2K.}\label{f20}
\end{figure}

The critical fields for the metamagnetic transitions are $H_c
\approx$  14kG, 31kG, 45kG for $H \| [120]$, 13kG, 31kG, 60kG for
$H \| [010]$, and 3kG, 12kG, 46kG, 50kG, 71kG for $H \| c$
respectively where it should be noted that for $H \| c \sim$ 3 kG
the transition is thought to be associated with domain rotation
(as discussed above). These values can also be seen in the $R(H)$
curves, shown in Fig. \ref{f20} for $H \| c$ and $H \perp c$,
except for the ones beyond 55kG, which was the upper field limit
for this magnetoresistance measurement. Once metamagnetism of
planar moments is better understood (see discussion section on
metamagnetism in TmAgGe below) the study of the angular dependence
of  metamagnetism in DyAgGe, a system with highly anisotropic but
non-planar, non-axial moments, should be interesting and hopefully
tractable.

\subsection{HoAgGe}
\begin{figure}
\begin{center}
\includegraphics[angle=0,width=120mm]{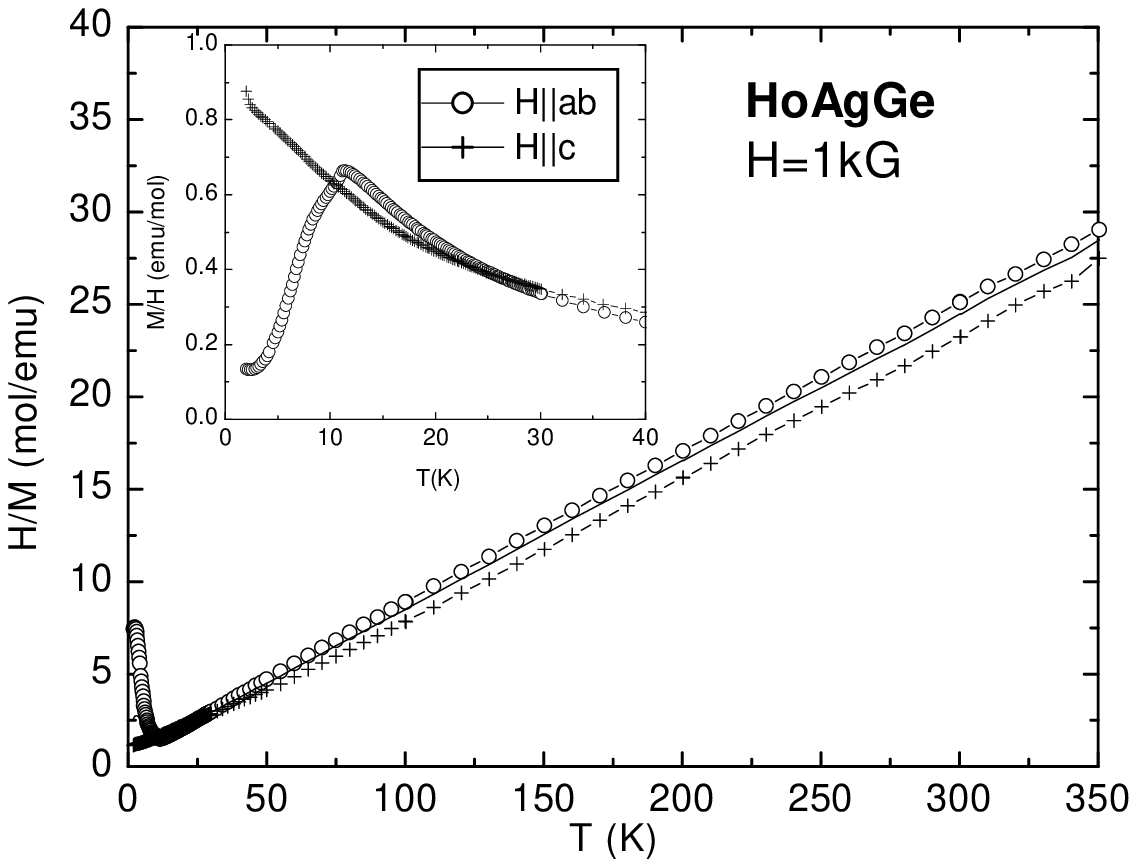}
\end{center}
\caption{Anisotropic inverse susceptibilities of HoAgGe and
calculated average (line); inset: low-temperature anisotropic
susceptibilities.}\label{f21}
\end{figure}

\begin{figure}
\begin{center}
\includegraphics[angle=0,width=120mm]{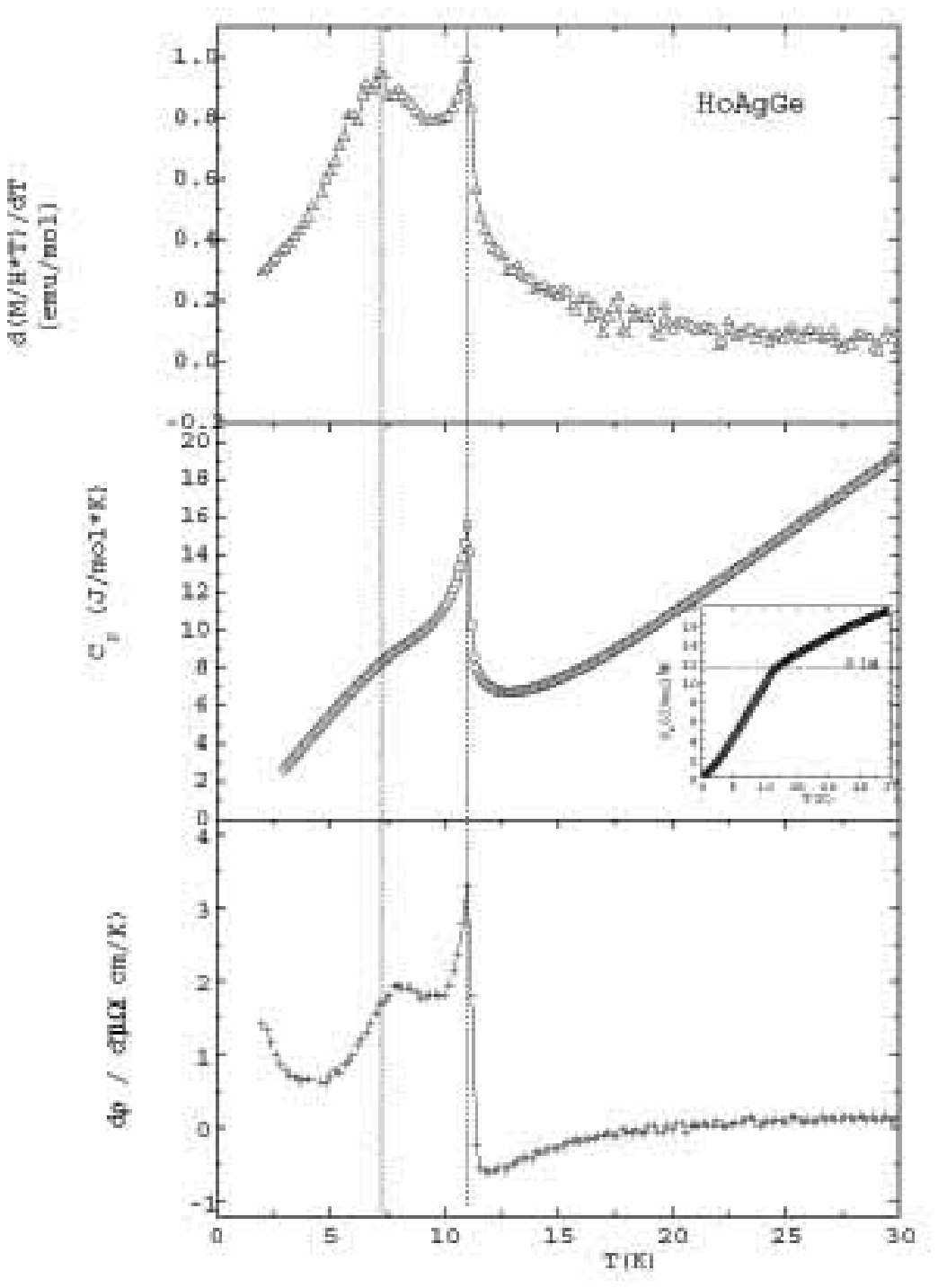}
\end{center}
\caption{(a) Low-temperature $d(\chi T)/dT$ of HoAgGe; (b)
Specific heat $C_p(T)$ with the magnetic entropy $S_m$ in the
inset; (c) low-temperature $d\rho/dT$; dotted lines mark the peak
positions as determined from (a).}\label{f22}
\end{figure}

The susceptibility of HoAgGe in the paramagnetic state is almost
isotropic, as can be seen in Fig. \ref{f21}. For temperatures
higher than 50K the susceptibility follows the Curie-Weiss law
$\chi(T)=C/(T+\Theta_P)$; the corresponding paramagnetic
temperatures $\Theta_P$ are given in Table \ref{t1}. From the
linear fit of the inverse susceptibility (for the polycrystalline
average) we get an effective moment $\mu_{eff}=10.0\mu_B$, close
to the theoretical value $\mu_{eff}($Ho$^{3+})=10.6\mu_B$. The
compound orders antiferromagnetically below N\'{e}el temperature
$T_N$=11.0K, whereas a spin reorientation transition occurs around
$T$=7.4K. Although $T_N$ shows up as a sharp, well-defined peak in
all three plots in Fig. \ref{f22}, the lower transition
temperature is indicated by broader peaks in $d(\chi T)/dT$ around
7.4K (Fig. \ref{f22}a), and in $d\rho/dT$ around 8.0K (Fig.
\ref{f22}c); specific heat plot shown in Fig. \ref{f22}b seems to
have an even broader feature close to these temperatures.  It
should be noted that, as in the case of TbAgGe, the measurements
on polycrystalline samples \cite{13} missed a transition ({\it
e.g.} the lower one for HoAgGe), whereas the reported temperature
of the upper transition falls close to our measured value.

\begin{figure}
\begin{center}
\includegraphics[angle=0,width=120mm]{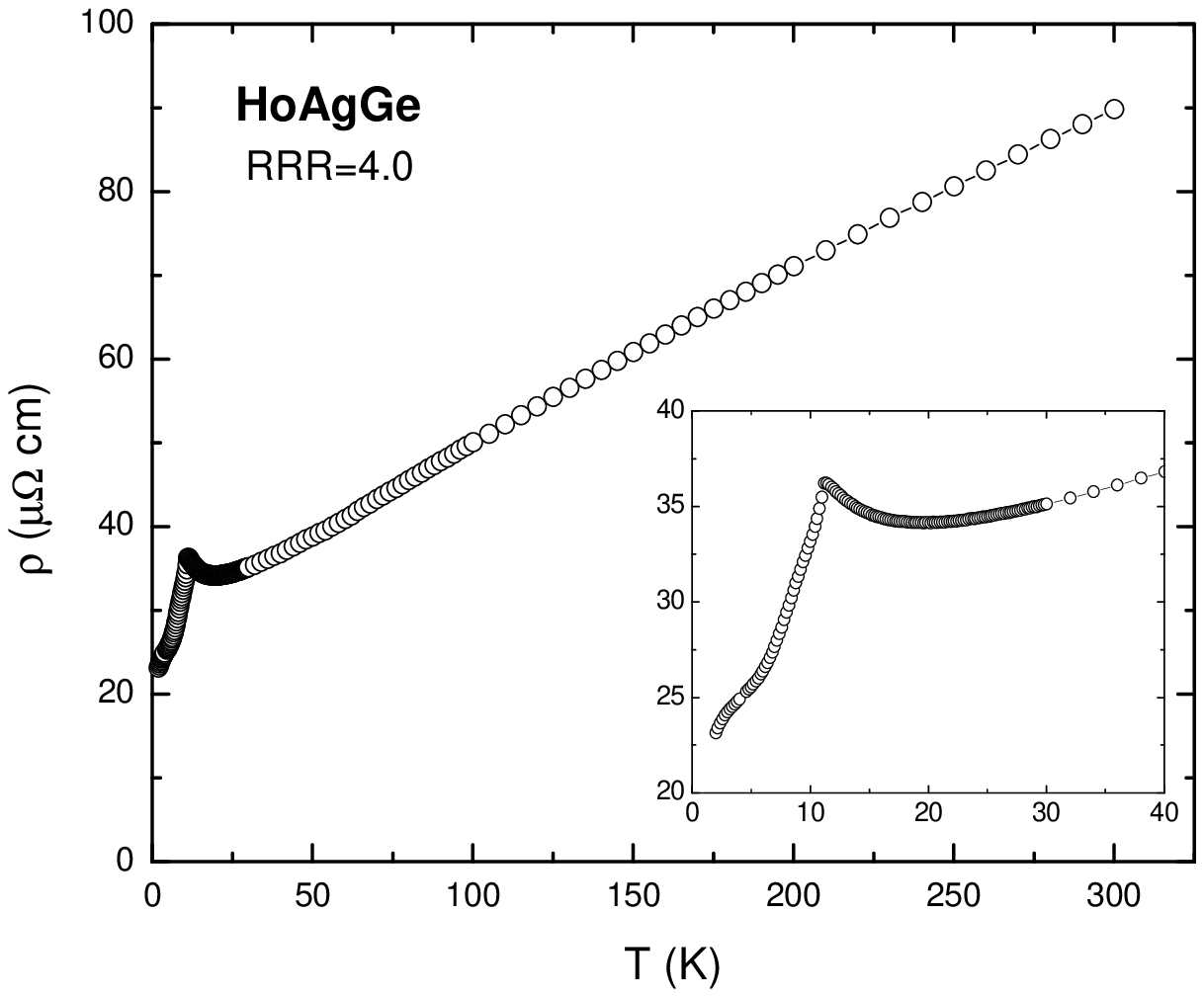}
\end{center}
\caption{Zero-filed resistivity of HoAgGe (inset: enlarged
low-temperature part).}\label{f23}
\end{figure}

The temperature dependent resistivity measurement (Fig. \ref{f23})
demonstrates the metallic character of this compound and the
residual resistivity ratio for HoAgGe is RRR=4.0. At a temperature
of about 19K a broad minimum in $\rho(T)$ is observed, and a drop
in resistivity due to the loss of spin disorder scattering in the
magnetically ordered phase.

From the specific heat measurements shown in Fig. \ref{f22}b we
calculated the magnetic entropy $S_m$ of this compound (shown in
the inset); the value of the entropy, $S_m \approx R \ln 4$, at
the change in slope that occurs around the ordering temperature
suggests that the magnetic order in HoAgGe emerges out of state
with a degeneracy of 4.

The magnetization curves as a function of applied field (Fig.
\ref{f24}a) reveal a series of metamagnetic transitions for $H$
applied in the $ab$-plane, with critical field values of around
11.0kG, 22.0kG and 35.0kG for $H \| [010]$, and 12.0kG, 23.0kG and
28.0kG respectively for $H \| [120]$, whereas when $H$ is applied
along the $c$-axis the magnetization curve can either be a broad
metamagnetic transition or a continuous spin-flop transition. The
theoretical value of the saturated magnetization expected for
Ho$^{3+}$ ($\mu_{sat}=10 \mu_B$) is not reached by fields up to
140kG in any of the three orientations; as mentioned before, this
can be due to the moments being along an inclined axis with
respect to the $c$-axis of the crystals, or further support the
idea of the spin-flop transition, especially given the continuous
increase in magnetization as $H$ is being increased. Another
possible explanation, which we detail for TmAgGe, is related to
the crystal structure of these compounds and will be discussed
below.

We have also measured magnetoresistance for $H \| ab$ (Fig.
\ref{f24}b), and it is consistent with the various metamagnetic
states seen in $M(H)$. Given the geometry of the crystals, there
was more uncertainty in orienting the resistance pieces than the
ones for magnetization measurements; consequently, we can infer
the approximate orientation of the field in the magnetoresistance
measurement by comparison of these data with the $M(H)$ curves for
$H \| ab$: since various features in $\Delta \rho(H)/\rho(0)$
occur close to the critical fields in $M([120])$ (Fig.
\ref{f24}b), it seems that $H$ was almost parallel to the $[120]$
direction.

\subsection{ErAgGe}
\begin{figure}
\begin{center}
\includegraphics[angle=0,width=120mm]{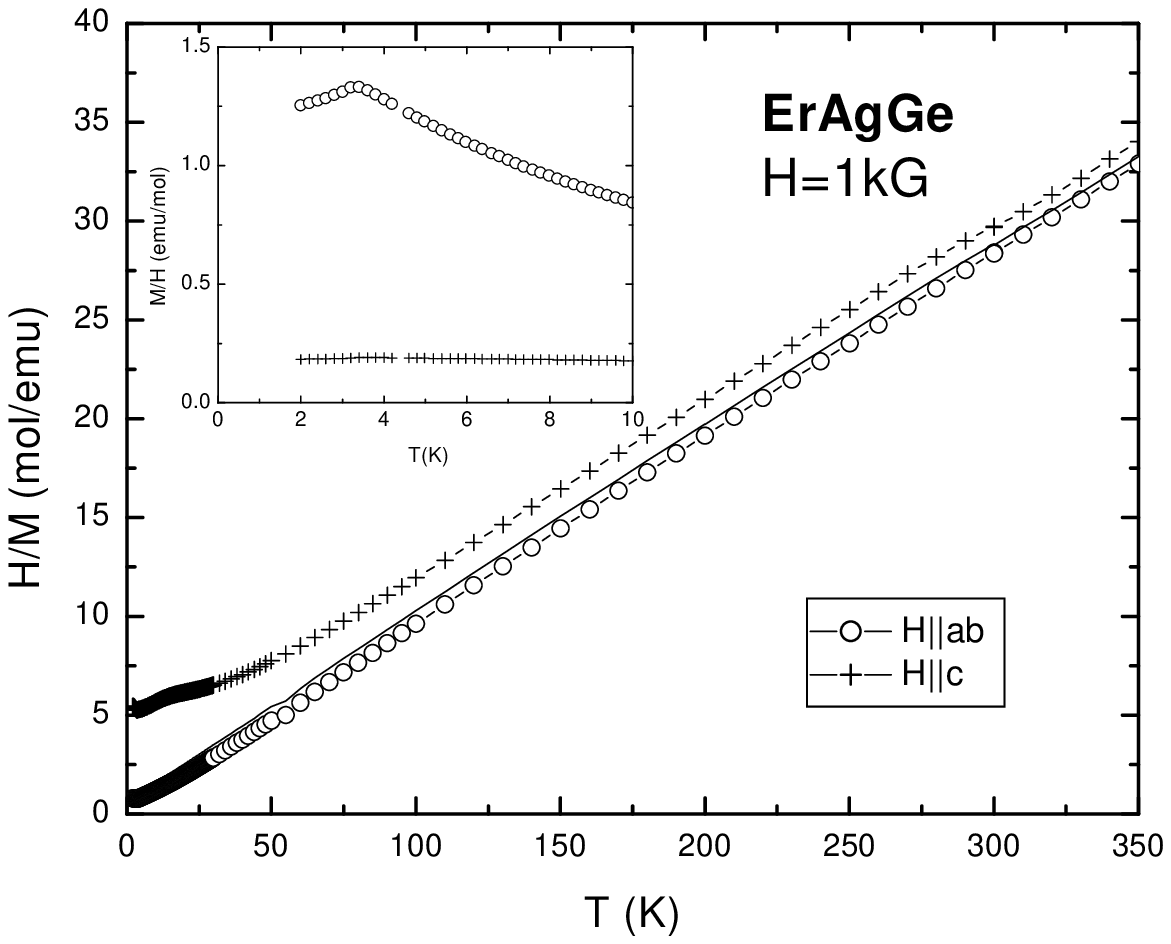}
\end{center}
\caption{Anisotropic inverse susceptibilities of ErAgGe and
calculated average (line); inset: low-temperature anisotropic
susceptibilities.}\label{f25}
\end{figure}

So far in the RAgGe series, we have seen a progression from axial
anisotropy in TbAgGe toward $M_{ab} \sim M_c$ for HoAgGe. ErAgGe
continues this trend with the local Er moments becoming far more
planar in nature (Fig. \ref{f25}). This is analogous to the trend
seen in many tetragonal systems \cite{1,2}, in which the change in
sign of the $B_{20}$ CEF parameter causes a switch from planar to
axial moments between Ho and Er \cite{20,21}.

The inverse susceptibility of ErAgGe is linear above 75K, with
$\mu_{eff}=9.3\mu_B$, fairly close to the theoretical value of
9.6$\mu_B$ for Er$^{3+}$. The anisotropic Weiss temperatures
$\Theta_P$ are given in Table \ref{t1}. For a 1kG field applied
parallel to the $ab$-plane, magnetic ordering is observed below 3K
(inset, Fig. \ref{f25}); a smaller feature seems to indicate the
same ordering temperature for field applied parallel to the
$c$-axis.

\begin{figure}
\begin{center}
\includegraphics[angle=0,width=120mm]{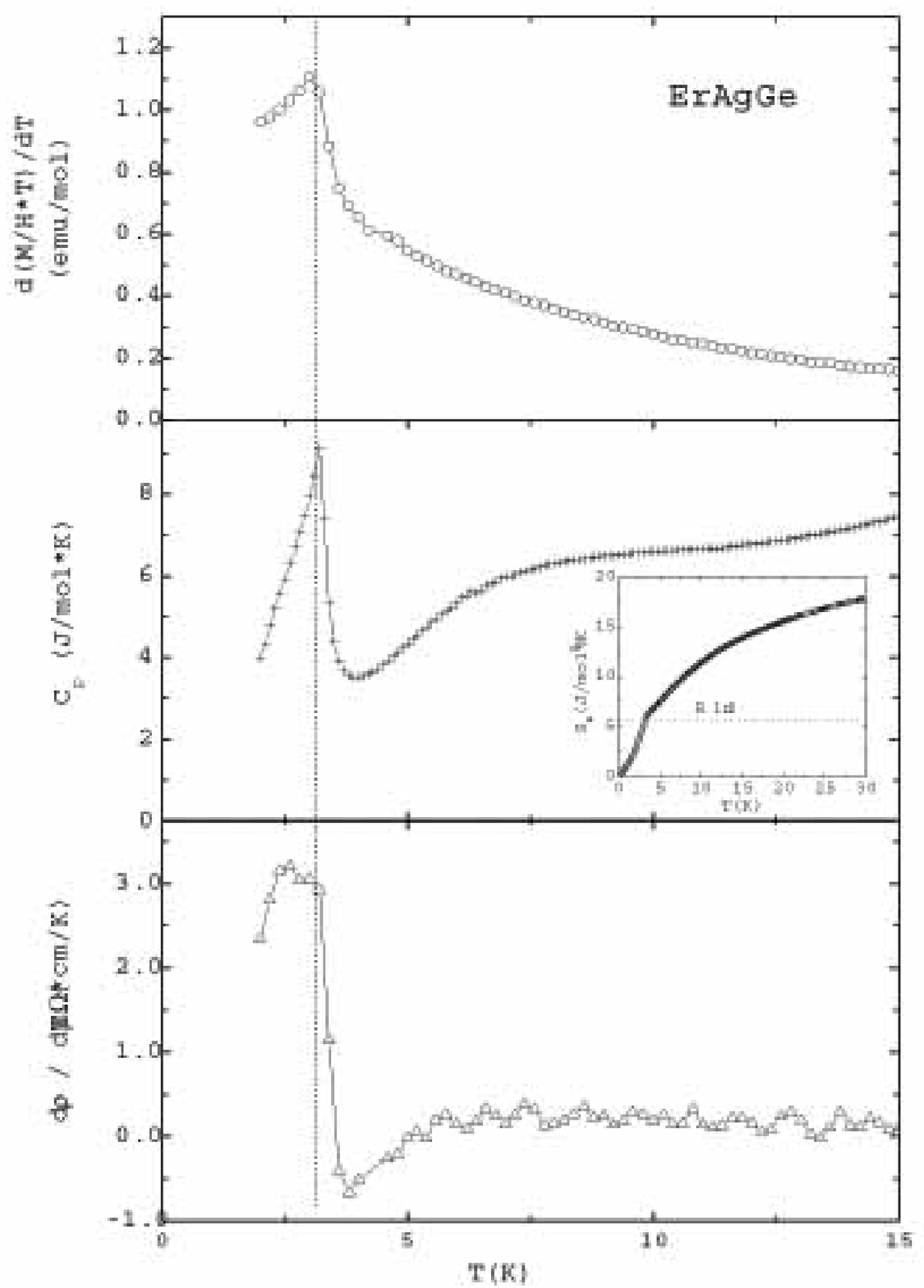}
\end{center}
\caption{(a) low-temperature $d(\chi T)/dT$ of ErAgGe ; (b)
Specific heat $C_p(T)$ with the magnetic entropy $S_m$ in the
inset; (c) low-temperature $d\rho/dT$; dotted line mark the peak
position as determined from (b).}\label{f26}
\end{figure}

\begin{figure}
\begin{center}
\includegraphics[angle=0,width=120mm]{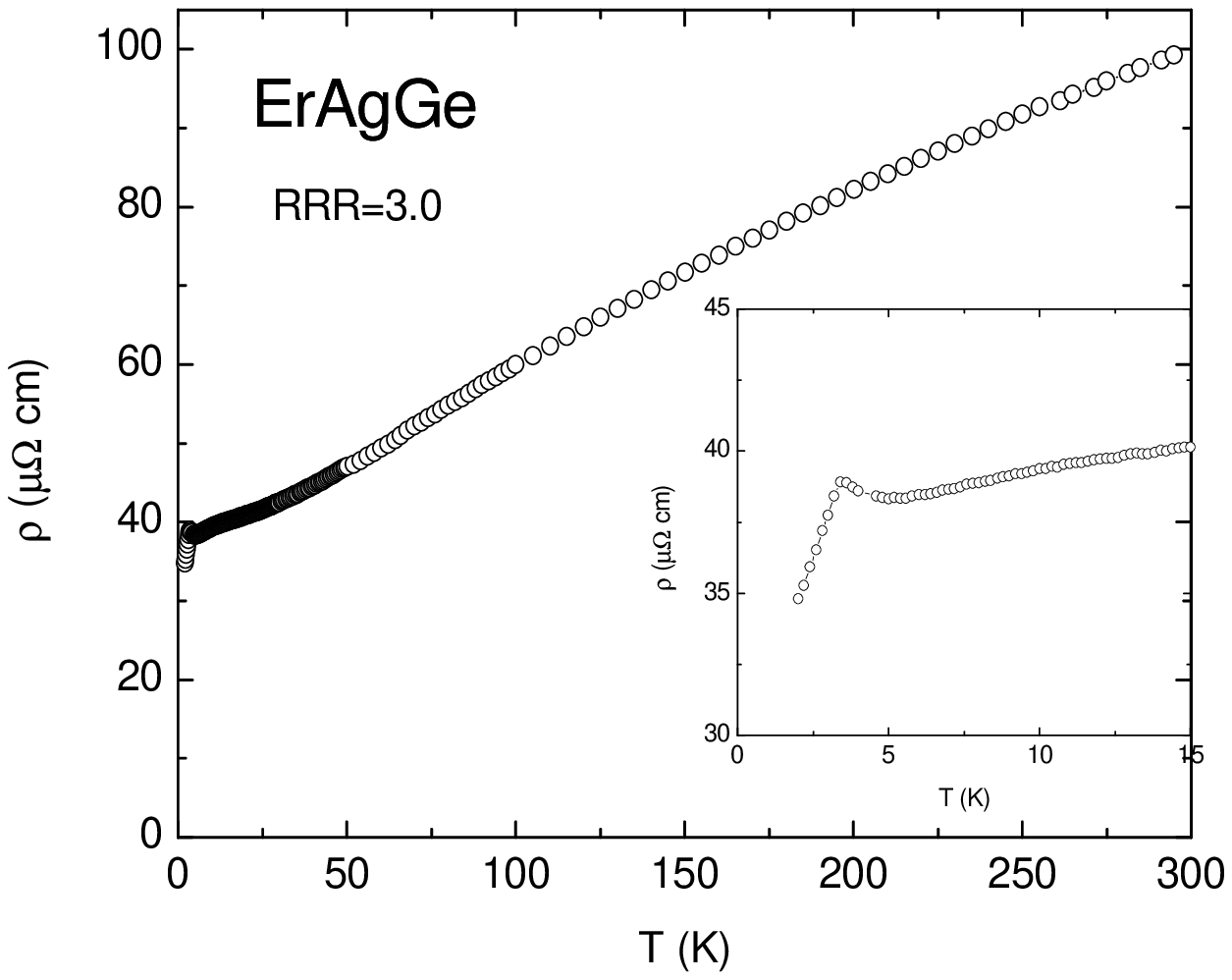}
\end{center}
\caption{Zero-field resistivity of ErAgGe (inset: enlarged
low-temperature part).}\label{f27}
\end{figure}

The temperature dependent resistivity is consistent with local
moment ordering and manifests a local maximum in $d\rho/dT$ at 3K
(Fig. \ref{f26}c). Below the ordering temperature a decrease in
resistivity (Fig. \ref{f27}) corresponds to the loss of
spin-disorder scattering, as the magnetic moments become
antiferromagnetically ordered.  The high temperature resistivity
is typical of intermetallic compounds, increasing up to $\sim 100
\mu\Omega cm$ at $T$=300K, and leading to a RRR value of
$\sim$3.0. The increasingly ubiquitous local minimum in the
resistivity can still be observed above $T_N$, centered near $T
\sim$ 6 K. In Fig. \ref{f26}b the specific heat shows a
well-defined peak at $T \approx$ 3.2K, very close to the
temperature of the maximum in $d(\chi T)/dT$ (Fig. \ref{f26}a).
The inset is a plot of the magnetic entropy of ErAgGe, where the
break in slope around $S_m \approx R \ln 2$ indicates that the
ground state of this compound is a Kramers doublet.

\begin{figure}
\begin{center}
\includegraphics[angle=0,width=120mm]{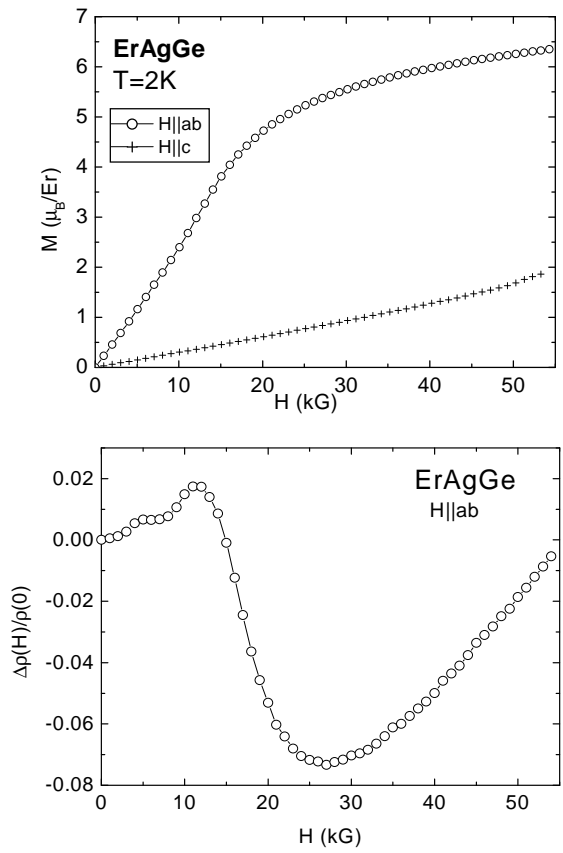}
\end{center}
\caption{(a) Anisotropic field-dependent magnetization for ErAgGe
at $T$=2K; (b) transverse magnetoresistance at $T$=2K.}\label{f28}
\end{figure}

In the $M(H)$ plots (Fig. \ref{f28}a) we again see clear
anisotropy, with the moments somewhat constrained to the basal
plane.  For field applied along the c direction, magnetization
linearly increases with field, up to $H \approx$ 50kG; right
before the maximum applied field $H$=55kG, an upturn in the $M(H)$
curve is apparent, possibly indicating a subsequent metamagnetic
transition. When field is applied parallel to the $ab$-plane, we
see a broad, poorly defined metamagnetic transition, as our
measurements are being taken at $T$=2K, close to $T_N$ of this
compound. The magnetoresistance for $H \| ab$ (Fig. \ref{f28}b)
shows a local maximum around $H$=11kG, which further indicates the
presence of a metamagnetic transition for this critical field.

\subsection{TmAgGe}
\begin{figure}
\begin{center}
\includegraphics[angle=0,width=120mm]{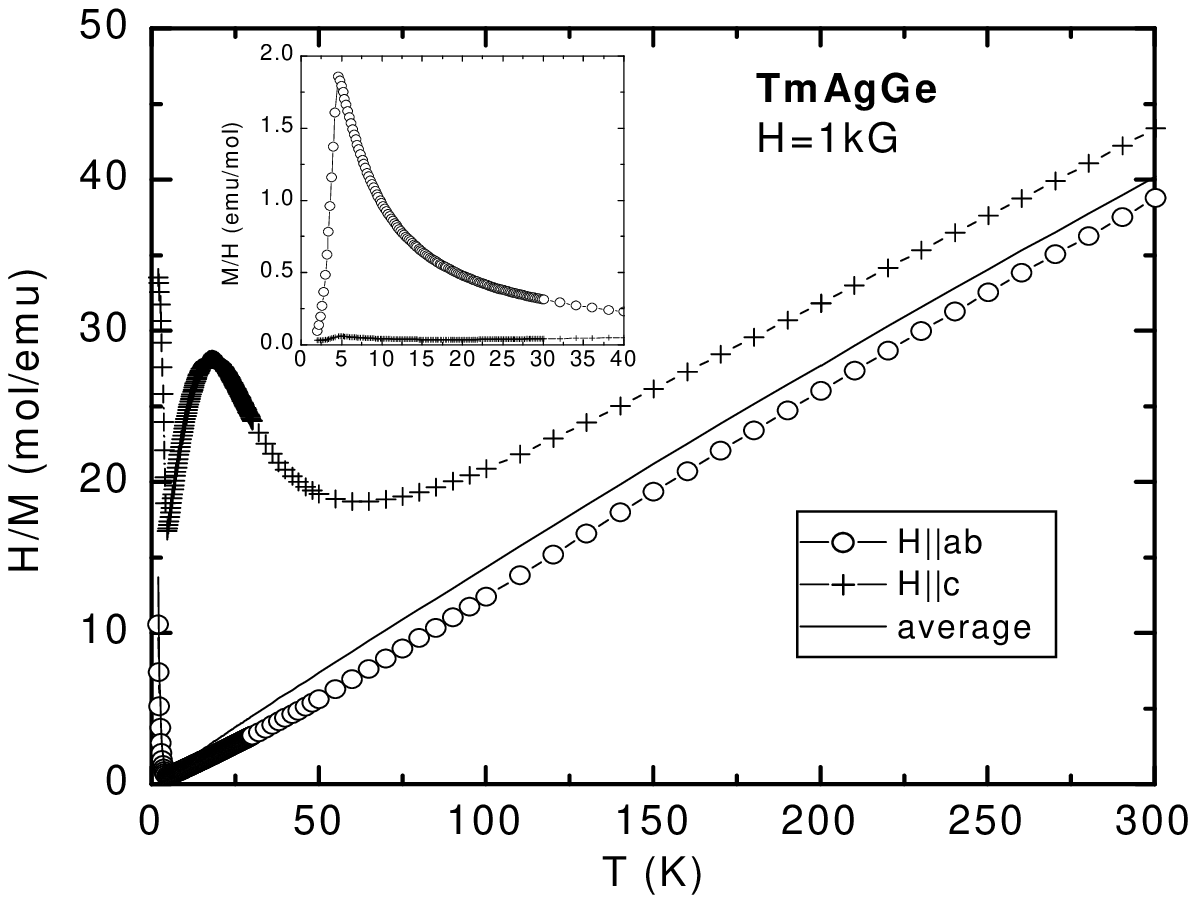}
\end{center}
\caption{Anisotropic inverse susceptibilities of TmAgGe and
calculated average (line); inset: low-temperature anisotropic
susceptibilities.}\label{f29}
\end{figure}

Magnetization measurements with field applied perpendicular and
parallel to the $c$-axis (Fig. \ref{f29}) indicate extreme
anisotropy in TmAgGe: $\chi_{ab}/\chi_c \approx 30$ at $T$=5.0K.
This temperature was chosen just above the antiferromagnetic
ordering temperature $T_N \approx$ 4.2 K. Above $\sim$100K,
inverse susceptibilities are linear with $\mu_{eff}=7.9\mu_B$,
close to the theoretical value of $\mu_{eff}($Tm$^{3+})=7.6\mu_B$.

\begin{figure}
\begin{center}
\includegraphics[angle=0,width=120mm]{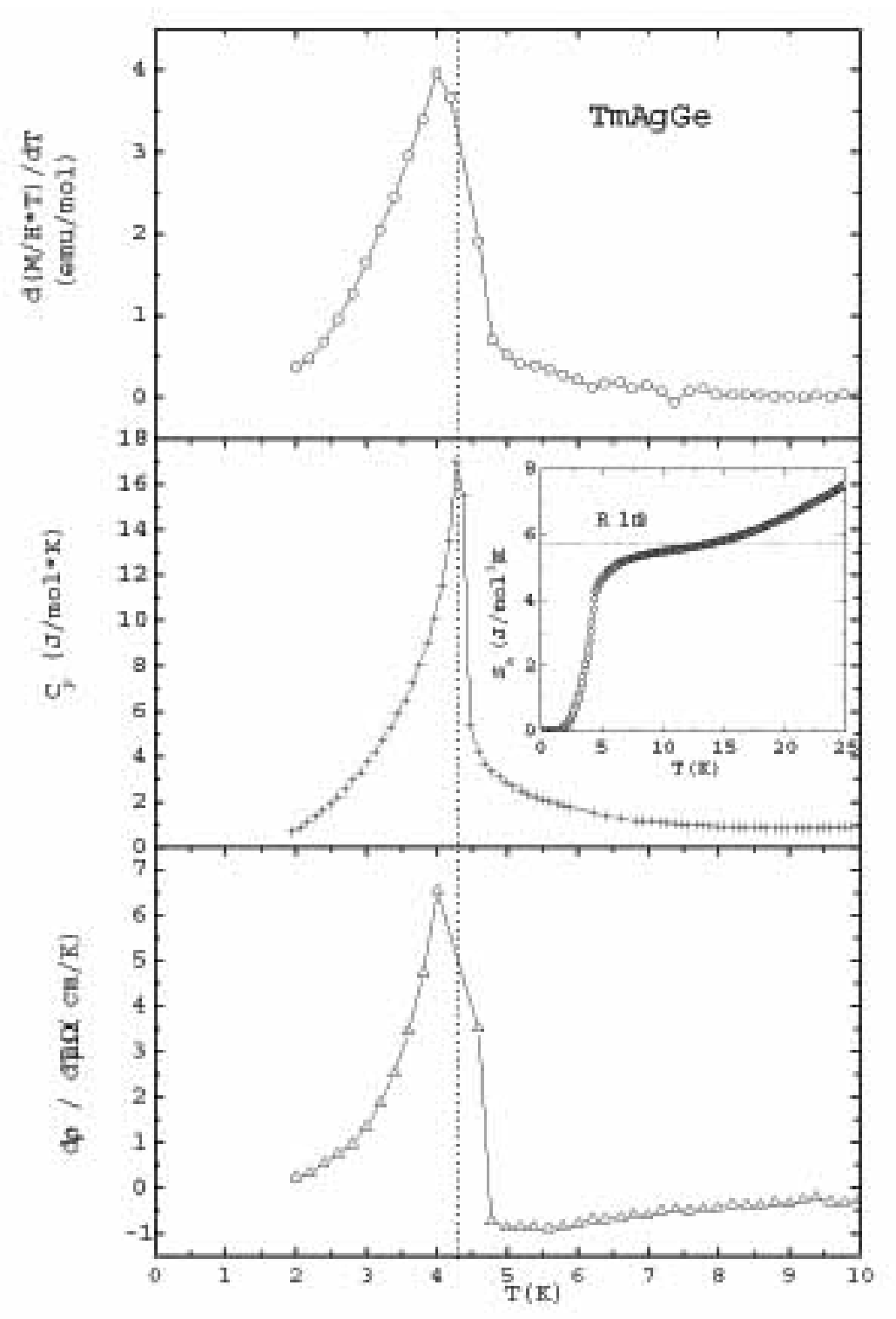}
\end{center}
\caption{(a )Low-temperature $d(M/H*T)/dT$ for TmAgGe; (b)
specific heat $C_p(T)$ with the magnetic entropy $S_m$ in the
inset; (c) low-temperature $d\rho/dT$; dotted line marks the peak
position as determined from (a).}\label{f30}
\end{figure}

\begin{figure}
\begin{center}
\includegraphics[angle=0,width=120mm]{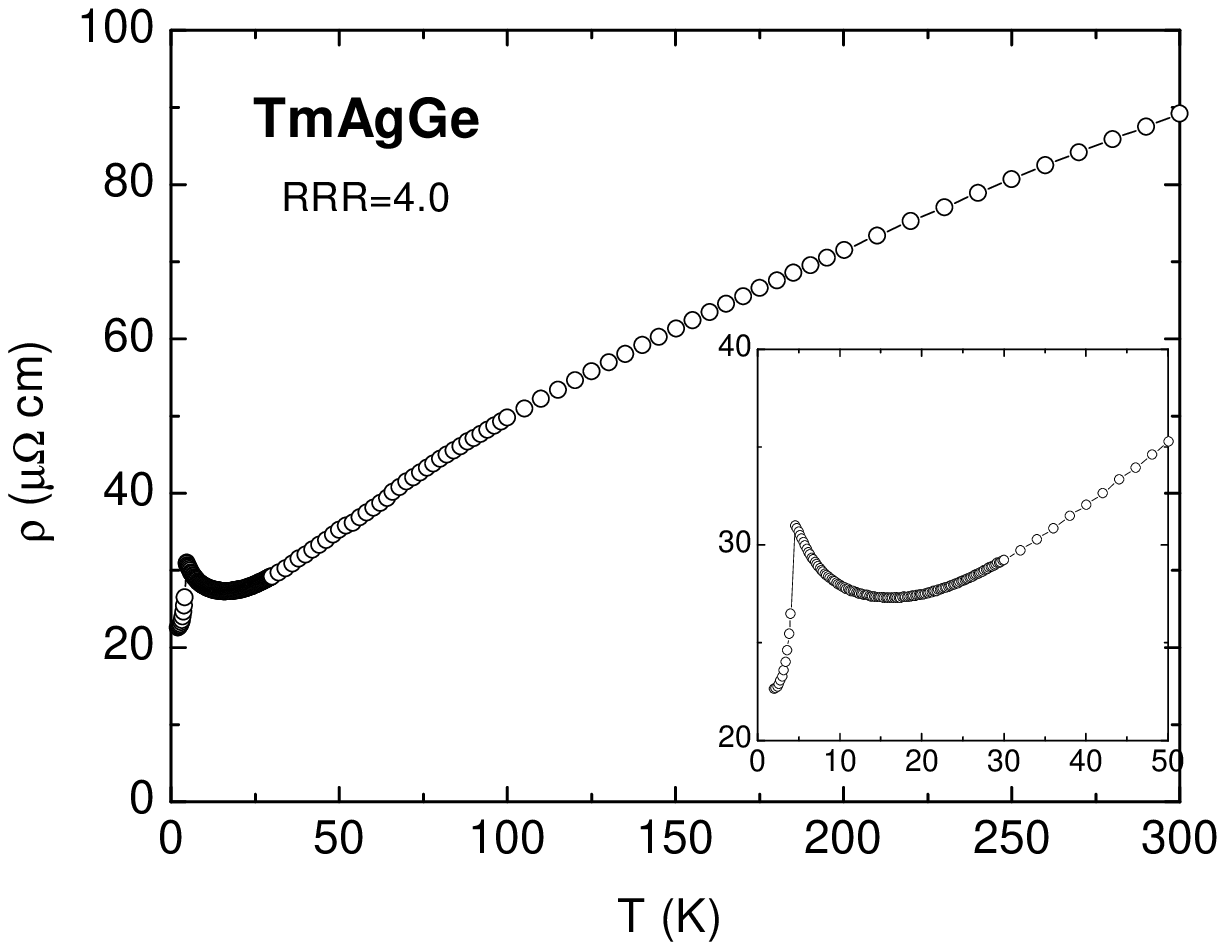}
\end{center}
\caption{Zero-field resistivity of TmAgGe (inset: enlarged
low-temperature part).}\label{f31}
\end{figure}

Figure \ref{f30} shows a sharp peak in $C_p(T)$ as well as $d(\chi
T)/dT$ for 4.2$\pm$0.1 K.  The zero-field resistivity data shown
in Fig. \ref{f31}, supports the ordering temperature inferred from
the thermodynamic data.  The plot of $d\rho/dT$ (Fig. \ref{f30}c)
has a clear peak between 4.0 and 4.6 K. In the inset of Fig.
\ref{f30}b, the calculated magnetic entropy $S_m(T)$ has a break
in slope close the transition temperature $T_N$, for an entropy
value of $\sim R\ln 2$, indicative of a doublet ground state or
two closely spaced singlets.  Below $T_N$, loss of spin disorder
scattering is apparent from the sudden drop in resistivity, due to
the antiferromagnetic ordering of the Tm$^{3+}$ moments. Above the
ordered state, the resistance goes through a broad local minimum
around 15K (far higher than $T_N$), after which it starts
increasing; for temperatures higher than $\sim$100K it becomes
approximately linear. The residual resistivity ratio RRR $\approx$
4.0 reflects acceptable quality of these crystals.

Clear metamagnetism is seen in the $M(H)$ plot in Fig. \ref{f32}a
for in-plane field orientation, as well as in magnetoresistance
measurements shown in Fig. \ref{f32}b. Below $H$=70kG, the
in-plane magnetization curves in Fig. \ref{f32}a show two
metamagnetic transitions for $H \| [120]$, with critical field
values $H_{c1} \approx$ 4.25kG and $H_{c2} \approx$ 9.25kG; for
the other orientation, $H \| [010]$, they merge into a single
transition with a critical field around $H_c \approx$ 6.0kG. A
very complex angular dependence of the critical fields, as well as
of the locally saturated magnetization values, can thus be
anticipated; this is subject of a separate, more detailed study
\cite{22}.

Similar to the case of HoAgGe, we compare the position of any
features revealed by the magnetoresistance measurement in Fig.
\ref{f32}b to the critical field values that we get from $M(H \|
ab)$ curves. Below 10kG, there is one obvious peak, and another
possible change of slope in the plot in Fig. \ref{f32}b; therefore
we could probably assume that the field was close to the $[120]$
direction, for which two metamagnetic transitions can be seen in
$M(H)$. However, there is another broad peak in the
magnetoresistance that can not be correlated with any feature in
magnetization, and which requires further investigation.

\subsection{YbAgGe}

\begin{figure}
\begin{center}
\includegraphics[angle=0,width=120mm]{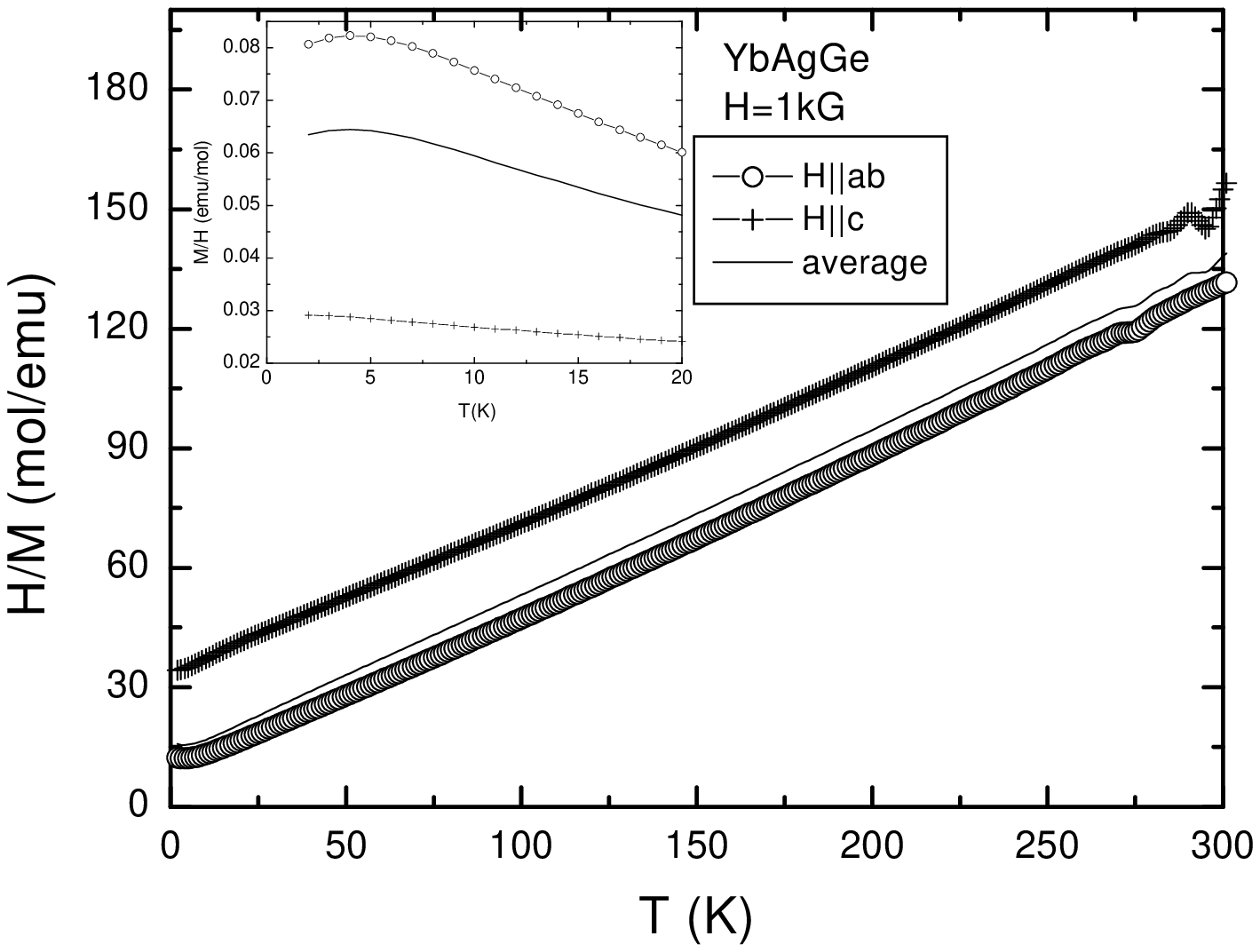}
\end{center}
\caption{Anisotropic inverse susceptibilities of YbAgGe and
calculated average (line); inset: low-temperature anisotropic
susceptibilities. }\label{f33}
\end{figure}

\begin{figure}
\begin{center}
\includegraphics[angle=0,width=120mm]{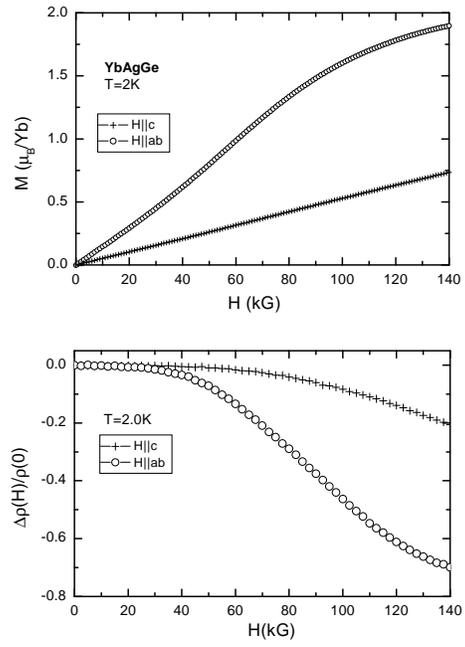}
\end{center}
\caption{(a) Anisotropic field-dependent magnetization and (b)
magnetoresistance for YbAgGe at $T$=2K.}\label{f34}
\end{figure}

YbAgGe is a compound with some distinctly different properties
compared to all previous members of the series. Fig. \ref{f33}
shows the inverse anisotropic susceptibility in an applied field
$H$=1kG. It is linear above $\sim$20K, indicating Curie-Weiss
behavior at high temperatures with an effective moment of $\sim
4.4 \mu_B/$Yb; however, below this temperature, as can be seen in
the inset from low temperature susceptibility, there is no sign of
magnetic ordering down to 1.85K. Instead there is an apparent loss
of local moment behavior, manifesting itself as a levelling off of
the susceptibility. Also, no distinct features appear in the
$M(H)$ or magnetoresistance data at $T$=2K (Fig. \ref{f34}a,b):
for $H \| c$ magnetization linearly increases with $H$ up to
$H$=140kG, whereas the in-plane data show that the compound is
probably approaching saturation (the fact that the high field
magnetization is lower than calculated
$\mu_{sat}($Yb$^{3+})=4.0\mu_B$ is possibly a result of the
crystal structure and CEF anisotropy of the RAgGe series, see the
discussion below); magnetoresistance at this temperature is
consistent with what one would expect for anisotropic,
paramagnetic metal.

\begin{figure}
\begin{center}
\includegraphics[angle=0,width=120mm]{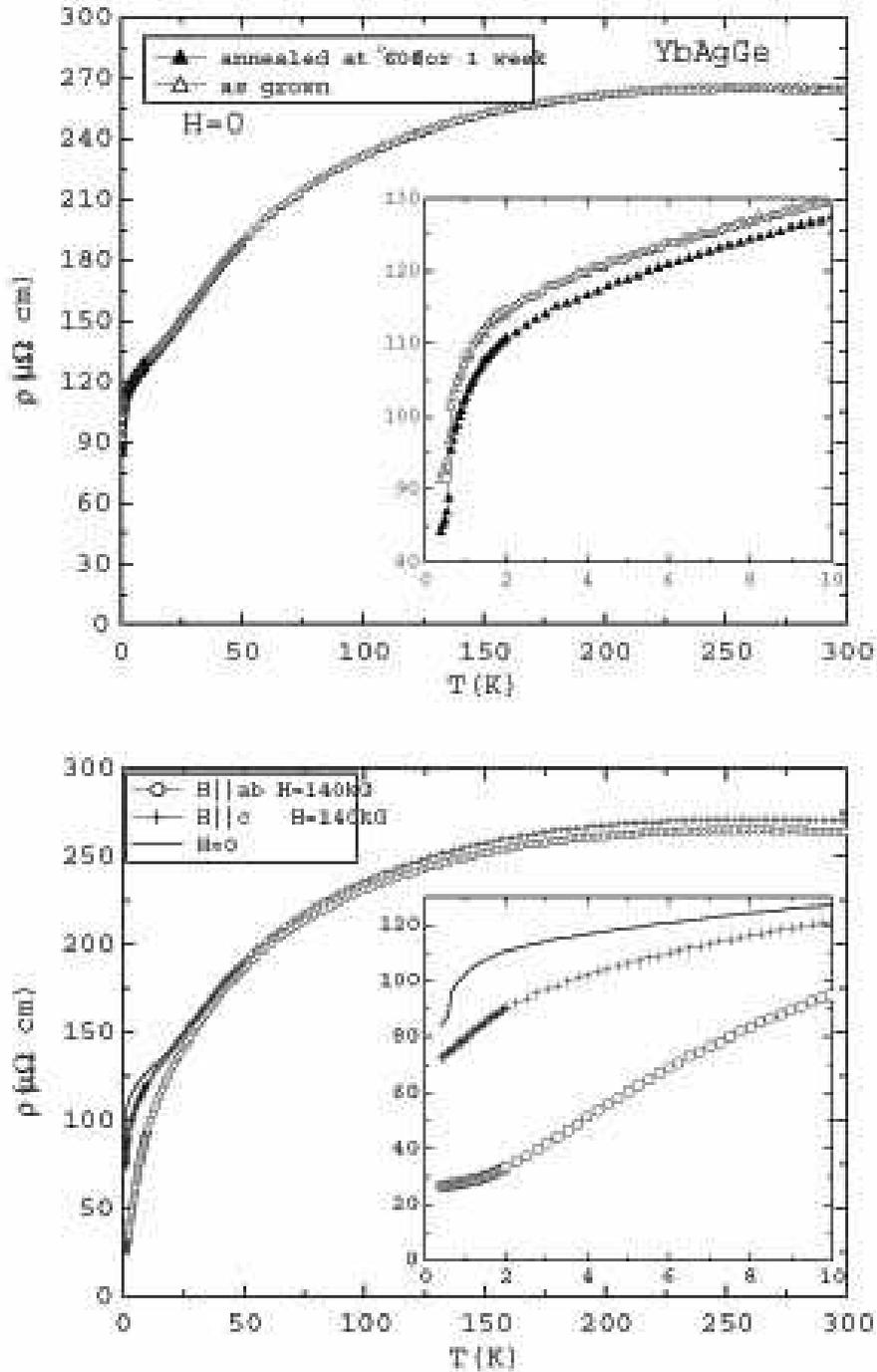}
\end{center}
\caption{(a) Zero-field resistivity of as grown and annealed
YbAgGe, (inset: enlarged low-temperature part); (b) anisotropic
transverse resistivity for annealed YbAgGe in applied field
$H$=140kG (small symbols) and for $H$=0 (line).}\label{f35}
\end{figure}

As can be seen in Fig. \ref{f35}a, no significant change occurred
after annealing (600$^\circ$C for 1 week), and the RRR increase
(from 2.8 up to 3.1) was slight.  The annealing temperature was
limited by the possibility of melting the small amount of residual
flux on the sample (note the Ag-Ge eutectic at
$\sim$650$^\circ$C). The temperature dependences of the
resistivity, and the specific heat (shown below) were almost
identical before and after the heat treatment of the crystals.

On the other hand the temperature dependent electrical resistivity
can be changed significantly by the application of magnetic field.
When large field is applied (Fig. \ref{f35}b), there is an
increase in the RRR value: for $H \| ab$, the RRR value increases
dramatically to RRR $\sim$ 10. This indicates that crystal purity
is quite high and apparently the large residual resistivity
$\rho_0$ for $H$=0 is due to sensitivity of hybridized Yb state to
relatively minor disorder.

\begin{figure}
\begin{center}
\includegraphics[angle=0,width=120mm]{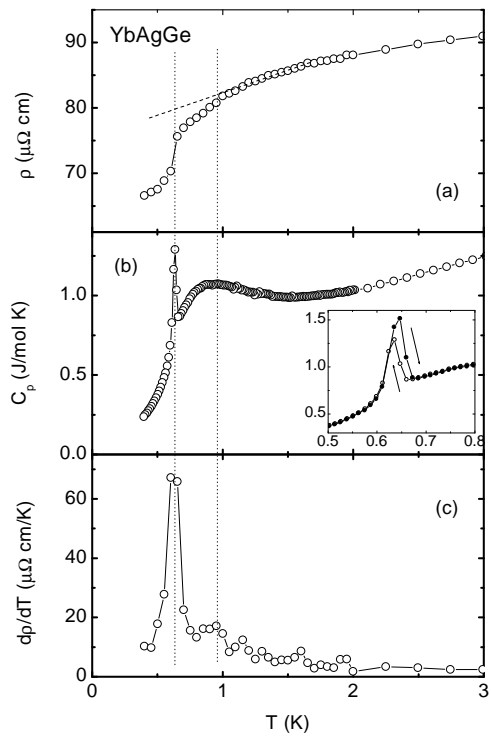}
\end{center}
\caption{Low temperature part of (a) resistivity; (b) specific
heat (inset: $C_p(T)$ on warming and cooling); (c) $d\rho /dT$.
Lines mark magnetic transitions.}\label{f36}
\end{figure}

\begin{figure}
\begin{center}
\includegraphics[angle=0,width=120mm]{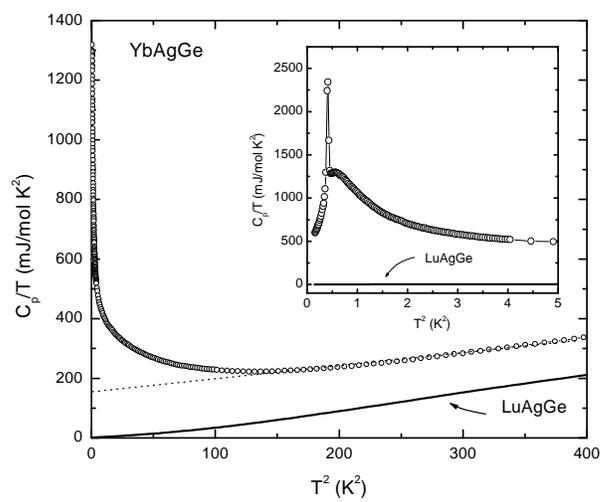}
\end{center}
\caption{$C_p(T^2) /T$ for YbAgGe (open circles) with a high
temperatures linear fit (dotted line) giving $\gamma$  value of
$(154.2\pm2.5)$ mJ/mol K$^2$; line: $C_p(T^2) /T$ for non-magnetic
LuAgGe. Inset: low temperature part.}\label{f37}
\end{figure}

In order to characterize this compound at lower temperatures we
have to rely on the resistivity (Figs. \ref{f35} and \ref{f36})
and specific heat (Figs. \ref{f36} and \ref{f37}) measurements
which were taken down to 0.4K. (Due to the limitations imposed by
our measurement systems magnetization data only go down to 1.85K.)
As the sample is cooled below 1.8 K there are two features visible
in both the resistivity and specific heat.  At 1.0 K there is a
slight but clear change in slope of the resistivity and there is a
relatively broad maximum in the specific heat.  At 0.65 K there is
an extremely sharp drop in resistivity and as well as a sharp peak
in the specific heat.  Specific heat data were taken upon heating
as well as cooling of the sample (Fig. \ref{f36}b, inset)
indicating that if there is any hysterisis associated with the
0.65 K transition it is smaller than the peak width.

\begin{figure}
\begin{center}
\includegraphics[angle=0,width=120mm]{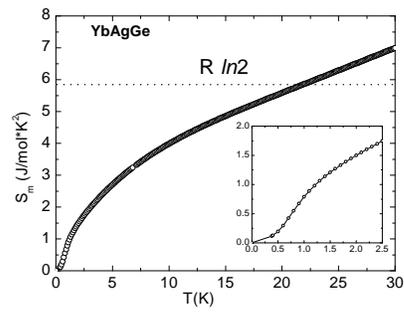}
\end{center}
\caption{Magnetic entropy for YbAgGe. Inset: low temperature
part.}\label{f38}
\end{figure}

The calculated magnetic entropy $S_m$ (Fig. \ref{f38}) at $T$ =
1.0 K is significantly less than $R \ln 2$.  Based on this we can
assume that the transition corresponds to small magnetic moment
ordering. This is in agreement with the enhanced electronic
specific heat coefficient $\gamma \ge 154$ mJ/mol K$^2$ that we
get from extrapolation of the high temperature part of the $C_p/T$
{\it vs.} $T^2$ (Fig. \ref{f37}) to $T^2$=0.

Low temperature small magnetic moment ordering is not the only
feature indicating that YbAgGe  is probably a heavy Fermion
compound:  deviations from the Curie-Weiss behavior in
$\chi_{ave}$ (for which CEF effects are cancelled to the first
order) below $\sim$20 K, and quite large Weiss temperature
($\Theta_{ave} \approx -30$ K) suggest that Yb $4f$ levels may be
significantly hybridized. If this is true, an estimate of the
Kondo temperature is given by $\Theta/10 \le T_K \le \Theta$
\cite{23}, i.e. 3K $\le T_K \le$ 30K.  Specific heat data plotted
as $C_p/T$ {\it vs.} $T^2$ (Fig. \ref{f37}) reveal a distinct
enhancement of the electronic specific heat coefficient (compared
with the data for LuAgGe on the same plot). This $\gamma \sim$ 150
mJ/mol K$^2$ value, already being significantly enhanced by
itself, is apparently a lower limit of $\gamma$ since a
significant upturn in $C_p/T$ {\it vs.} $T^2$ is observed below
approximately 10K, similar to that seen in many heavy fermion
compounds \cite{24}, including for example YbNi$_2$B$_2$C
\cite{25} and YbRh$_2$Si$_2$ \cite{26}.  Keeping in mind that
YbAgGe has a low temperature magnetic ordering, and that therefore
an ambiguity in the evaluation of low temperature electronic
specific heat coefficient is present, Fig. \ref{f37} suggests that
low temperature $\gamma$ is within the range of $\sim$150 mJ/mol
K$^2$ to 1 J/mol K$^2$. This crude estimate allows us to classify
YbAgGe as a new Yb heavy fermion compound, with low temperature
reduced moment magnetic ordering, which we already reported in
\cite{27}. Using the single impurity relation \cite{24}, we can
estimate the Kondo temperature as $T_K=w_N\pi^3R/6\gamma$, where
$w_N$=0.4107 is the Wilson number and $R$ is the gas constant.
Using the aforementioned range of the value of $\gamma$, the Kondo
temperature can be evaluated as 15 K $< T_K <$ 120 K.
Additionally, we can estimate the Wilson ratio, $\mathcal{R} =
4\chi\pi^2k_B^2/3\gamma\mu_{eff}^2$  \cite{24} for YbAgGe using
$\chi$ and $\gamma$ determined at $T$ = 1.8 K, and the high
temperature effective moment $\mu_{eff}$, as $\mathcal{R} \approx
1.8$, which is close to $\mathcal{R} = 2$ expected for heavy
fermion compounds, and much higher than $\mathcal{R} = 1$, the
Wilson ratio for non-interacting electrons \cite{24}.

Having a N\'{e}el temperature, apparently associated with small
moment ordering, so close to $T$=0K makes YbAgGe an interesting
system for the study of the competition between magnetically
ordered and correlated ground state. It is anticipated that
pressure should be a possible parameter for stabilizing the
antiferromagnetic ground state.

While preparing this manuscript for publication, a very recent
report on the heavy fermion character in YbAgGe came to our
attention \cite{28}.  Although these authors did not go to low
enough temperatures to detect the ordering below 1 K, their other
data is in good agreement with what we have presented here and
elsewhere \cite{27}.

\section{Discussion}
Among many properties that we see in the RAgGe series, anisotropy
and metamagnetism are particularly interesting, specifically in
light of the crystal structure of these compounds:  a hexagonal
unit cell with a single rare earth site of orthorhombic point
symmetry.  Across the series, the magnetization is anisotropic,
going from axial (in TbAgGe) to extreme planar (in TmAgGe). The
magnetic and transport properties throughout the RAgGe series
proved to be anisotropic, due primarily to the CEF splitting of
the Hund's rule ground state multiplet. In Table \ref{t1} the
Weiss paramagnetic temperatures are given, for the two
orientations of the field, as well as for the polycrystalline
average. Negative values for $\Theta_{ave}$ for all R suggest
antiferromagnetic interactions between magnetic moments, although
Dy may be an exception, given the presence of the small
ferromagnetic component of the magnetization. TbAgGe has an easy
axis parallel to $c$ in the paramagnetic state. This is followed
by a progression towards a more isotropic case (R=Ho) while having
$\Theta_{ab}<\Theta_c$, whereas for R=Er-Yb the easy axis lies in
the $ab$-plane and $\Theta_{ab}>\Theta_c$. The analysis of this
anisotropy should allow for the determination of the leading term
in the crystal field Hamiltonian, similar to the case of
tetragonal systems. But a more complex calculation is needed for
our hexagonal compounds, where the R ions are located at sites
with {\it orthorhombic} point symmetry, and this is beyond the
scope of the current paper.

\begin{table}
\caption{Magnetic ordering temperatures, $T_m$, effective magnetic
moments and anisotropic paramagnetic Weiss temperatures
$\Theta_p$.} \label{t1}
\begin{tabular}{|c|c|c|c|c|c|c|} \hline
 & Tb & Dy & Ho & Er & Tm & Yb\\
\hline & 28.4$\pm$0.1, & 14.5$\pm$0.1, & 11.0$\pm$0.2, & & &
0.95$\pm$0.05,\\ $T_m$(K) & 24.6$\pm$0.1, & 12.15$\pm$0.1 &
7.7$\pm$0.1 & 3.1$\pm$0.1 & 4.15$\pm$0.1 & 0.65$\pm$0.05
\\& 19.8$\pm$0.1 & & & & & \\
\hline $\mu_{eff}(\mu_B/$R) & 9.7 & 10.4 & 10.0 & 9.3 & 7.9 &
4.4 \\
\hline $\Theta_{ab}$(K) & -53.8 & -25.5 & -10.1 & -3.6 & 7.5 &
-15.1 \\
\hline $\Theta_{c}$(K) & 1.0 & 7.4 & -1.9 & -36.4 & -76.3 &
-83.5 \\
\hline $\Theta_{ave}$(K) & -28.3 & -10.5 & -7.1 & -10.9 & -14.4 &
-30.1 \\
\hline
\end{tabular}
\end{table}

\begin{figure}
\begin{center}
\includegraphics[angle=0,width=120mm]{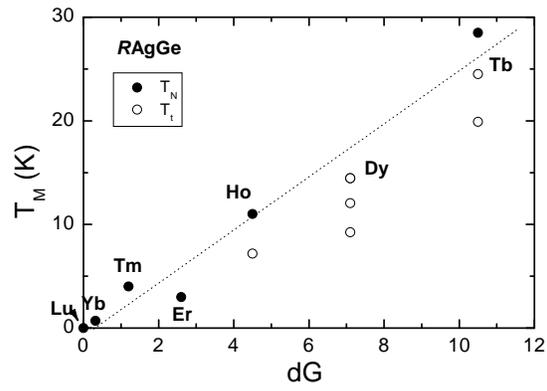}
\end{center}
\caption{Changes of the magnetic ordering temperatures $T_m$ for
R=Tb-Lu in the RAgGe series, with the de Gennes scaling parameter
dG (the dotted line represents the expected linear
dependence).}\label{f39}
\end{figure}

The ordering temperatures in RAgGe (R=Tb-Tm) approximately scale
with the de Gennes  factor $dG = (g_J - 1)^2J(J+1)$ (Fig.
\ref{f39}), where $g_J$ is the Land\'{e} $g$ factor and $J$ is the
total angular momentum of the R$^{3+}$ ion Hund's rule ground
state. This is consistent with the coupling between the conduction
electrons and the local magnetic moments giving rise to the long
range magnetic order via the RKKY interaction. However,
significant deviations from the linearity may be noticed,
suggesting other factors may be involved (i.e. strong CEF effects
constraining the moments to either the $ab$ plane or the $c$ axis,
as seen in the already mentioned tetragonal compounds
\cite{1,2,29}).

\begin{figure}
\begin{center}
\includegraphics[angle=0,width=120mm]{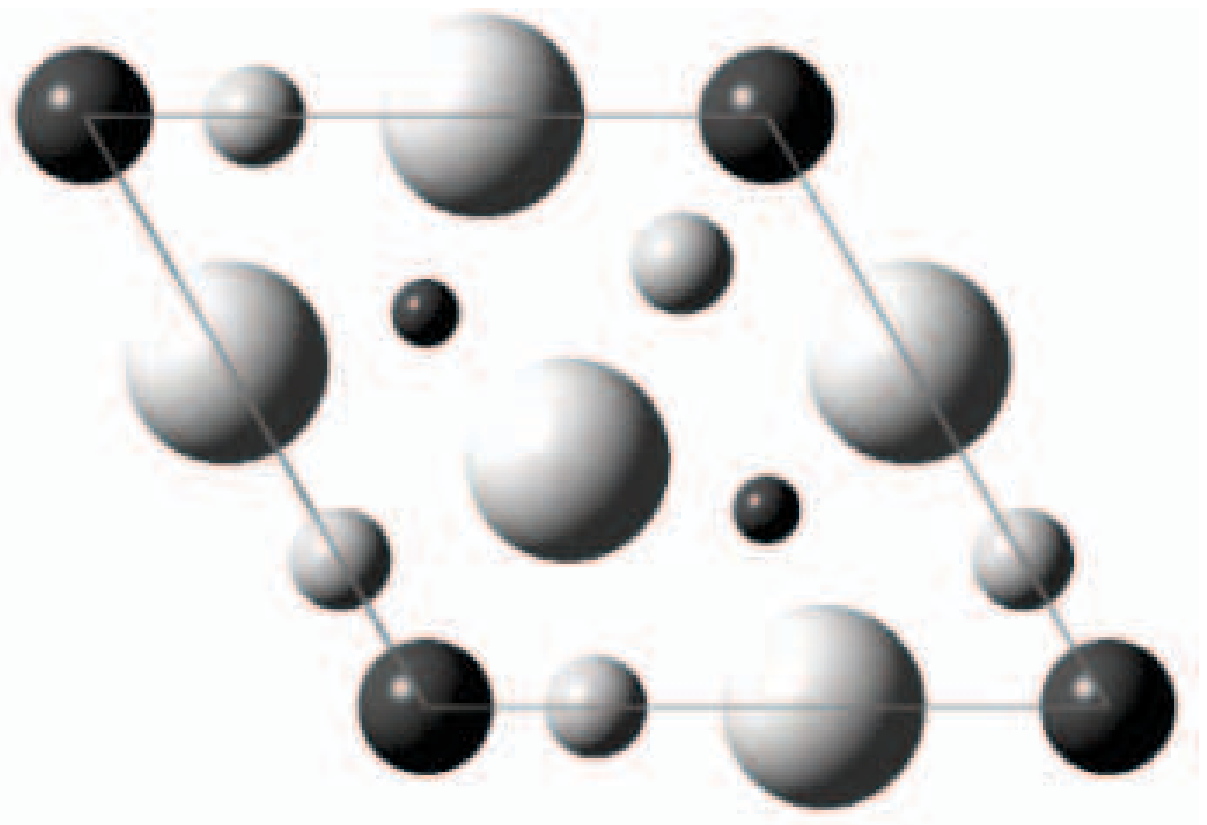}
\end{center}
\caption{$c$-axis projection of the RAgGe crystal structure. R -
light large circles and Ge - dark medium circles form R$_3$Ge
layers; Ag - smaller light and Ge - small dark circles form
Ag$_3$Ge$_2$ layer.}\label{f1}
\end{figure}

\begin{figure}
\begin{center}
\includegraphics[angle=0,width=120mm]{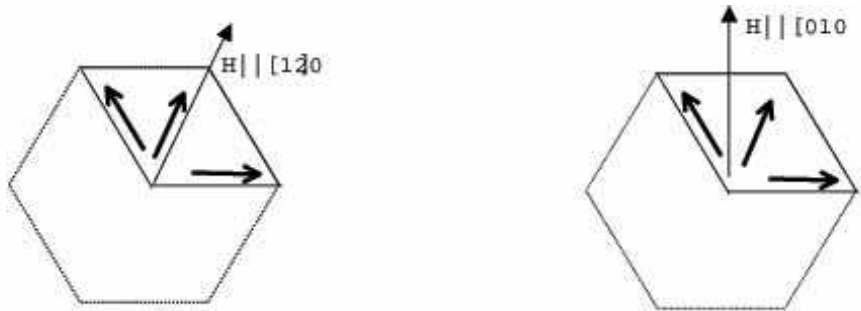}
\end{center}
\caption{Hexagonal unit cell with the magnetic moments (short
arrows) in the saturated state of TmAgGe; long arrows indicate the
field along the two high-symmetry directions of the crystal.
}\label{f40}
\end{figure}

Because the strong CEF splitting confines the magnetic moments to
the basal plane in TmAgGe this compound is the simplest candidate
in this series for a study of the metamagnetism.  Tm$^{3+}$ ions
occupy $3g$  Wyckoff sites, with $m 2 m$ ({\it orthorhombic})
point symmetry, leading to a hexagonal structure with three R ions
per unit cell (Fig. \ref{f1}). Consequently, referring to a single
unit cell, we can assume that the magnetic moments of the three R
ions behave like Ising systems, rotated by 120$^\circ$ with
respect to each other in the basal plane.  The saturated state is
reached when the three magnetic moment vectors add constructively
(Fig. \ref{f40}).

\begin{figure}
\begin{center}
\includegraphics[angle=0,width=120mm]{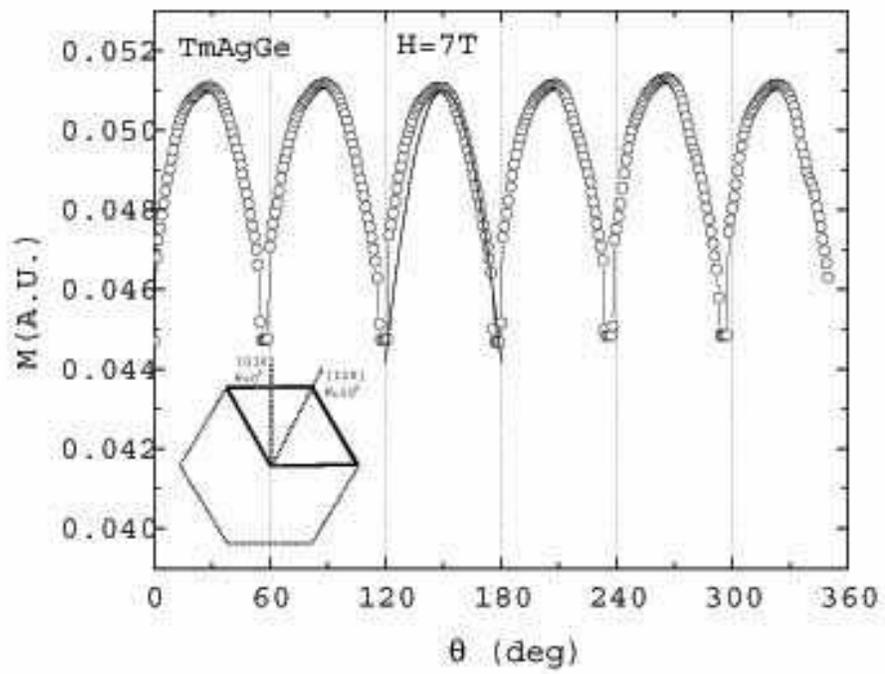}
\end{center}
\caption{Angular dependent magnetization in TmAgGe, for $H$=70kG
applied in the $ab$-plane (the solid line shown for
$\theta=120-180^\circ$ is $M(\theta)\sim \cos(\theta-30^\circ)$,
as expected from model).}\label{f41}
\end{figure}

To further examine the extreme basal plane anisotropy, we measured
$M(\theta)$ for an applied field $H$=70kG parallel to the $ab$
plane (Fig. \ref{f41}), where $\theta$ is the angle between the
direction of the field and one of the high symmetry directions.
Not only is the magnetization angular dependent, but also it
displays a {\it 6-fold} symmetry, consistent with the hexagonal
unit cell. The maxima in the angular dependent magnetization occur
for $\theta=(2n+1)*30^\circ$ ($n$-integer), when angle $\theta$ is
measured from the $[010]$ orthogonal axis. As a consequence,
within our model with a superposition of three Ising systems, for
the saturated state we can assume the magnetic moments
configuration shown in Fig. \ref{f40}.

Using Fig. \ref{f40}, we can evaluate the expected values of the
longitudinal magnetization (the quantity we measure) in the
saturated state, for two different orientations of the applied
field: when $H \| [120]$,

\begin{eqnarray}
M(H \| [120])&=&\frac{1}{3}[\mu_{sat}(Tm^{3+}) +
2\mu_{sat}(Tm^{3+}) \cos 60^\circ]= \nonumber \\
&=&\frac{1}{3}[\mu_{sat}(Tm^{3+})
+2\mu_{sat}(Tm^{3+})*\frac{1}{2}]=\nonumber \\
&=&\frac{2}{3}\mu_{sat}(Tm^{3+})=\frac{2}{3}* 7.0 \mu_B =
4.67\mu_B\nonumber
\end{eqnarray}

For the other {\it in-plane}, orthogonal direction ($H \| [010]$),
we get

\begin{eqnarray}
M(H \| [010])&=&\frac{1}{3}[0 +
2\mu_{sat}(Tm^{3+}) \cos 30^\circ]= \nonumber \\
&=&\frac{2}{3}\mu_{sat}(Tm^{3+}) \cos 30^\circ =\nonumber \\
&=&\frac{2}{3}\frac{\sqrt{3}}{2}\mu_{sat}(Tm^{3+})=\frac{2}{3}\frac{\sqrt{3}}{2}
7.0 \mu_B = 4.04\mu_B\nonumber
\end{eqnarray}

This results in a calculated ratio

\begin{eqnarray}
\frac{M(H\|[010])}{M(H\|[120])}\bigg|_{calc} = \cos 30^\circ =
\frac{\sqrt{3}}{2} = \frac{4.04}{4.67} = 0.867 \nonumber
\end{eqnarray}

If we compare these model calculations with the two corresponding
measured values from $M(\theta)$ in Fig. \ref{f41}, we notice that

\begin{eqnarray}
\frac{M(H\|[010])}{M(H\|[120])}\bigg|_{exp} = \frac{M(\theta =
0^\circ)}{M(\theta = 30^\circ)} = \frac{0.0447}{0.0551} = 0.875
\nonumber
\end{eqnarray}

is close to the anticipated value.  Together with the
experimental, angular dependent magnetization (open circles), Fig.
\ref{f41} also shows the angular dependence of $M$ (solid line),
calculated based on the above model for $\theta = 120 -
180^\circ$. The small, yet noticeable, deviations of the cosine
function from the measured data could have been caused by slight
misalignment of the extremely small sample used for the rotation
measurements; as a result, the rotation axis makes a small angle
with the $ab$-plane, resulting in a weak 2-fold modulation of the
of $M(\theta)$, yielding the asymmetric experimental peaks.

\begin{figure}
\begin{center}
\includegraphics[angle=0,width=120mm]{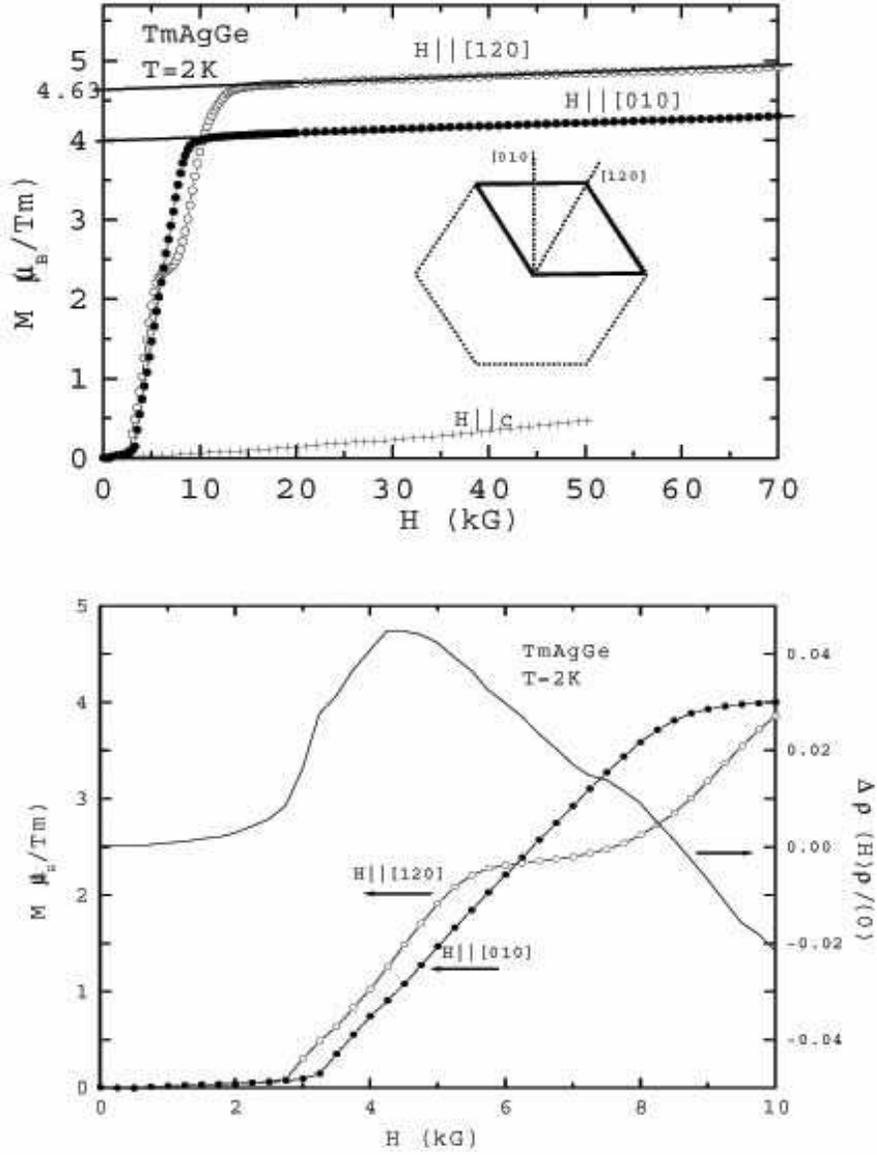}
\end{center}
\caption{(a) Anisotropic magnetization curves in TmAgGe, shown for
three orientations of the applied field; (b) Transverse
magnetoresistance  and $M(H$) curves for $H\|ab$, at
$T$=2K.}\label{f32}
\end{figure}

In Fig. \ref{f32}, above $\sim$15kG we see a wide plateau in
magnetization for the two high symmetry  orientations; the
corresponding magnetization values, even at highest applied field,
$H$=70kG, are $M(H\|[120])=4.92\mu_B$, and
$M(H\|[010])=4.30\mu_B$, far below $\mu_{sat}($Tm$^{3+})=7\mu_B$.
Instead of attributing this to additional metamagnetic transitions
beyond 70kG, we notice that the magnetization values for the
corresponding orientations of the field, and their ratio:

\begin{eqnarray}
\frac{M(H\|[010])}{M(H\|[120])}\bigg|_{exp} = \frac{4.30}{4.92} =
0.875 \nonumber
\end{eqnarray}

are consistent with our model. As the above ratio is close to the
expected $\cos30^\circ$ value, the absolute values of the
magnetization are 4.30$\mu_B$ and 4.92$\mu_B$, both larger than
the corresponding values within our model (i.e. 4.04$\mu_B$, and
4.67$\mu_B$ respectively). However, the extrapolations of the
high-field plateaus down to $H$=0 (straight lines in Fig.
\ref{f32}a) intersect the magnetization axis at 4.0$\mu_B$ and
4.63$\mu_B$, very close to the anticipated theoretical values.
This could be further indication of the validity of our model
applied to this compound, while the slow increase in magnetization
above alleged saturation could indicate a slow approach of the CEF
splitting energy.

The model in which we assume three Ising systems $120^\circ$ apart
in the $ab$-plane of a hexagonal unit cell, seems to describe well
the metamagnetism in TmAgGe; furthermore, it can be used to
describe at least one more compound isotypic to TmAgGe ({\it i.e.}
TbPtIn), and it will be used for a study of the angular dependent
metamagnetism in the two systems in a detailed paper yet to come
\cite{22}.

\begin{figure}
\begin{center}
\includegraphics[angle=0,width=120mm]{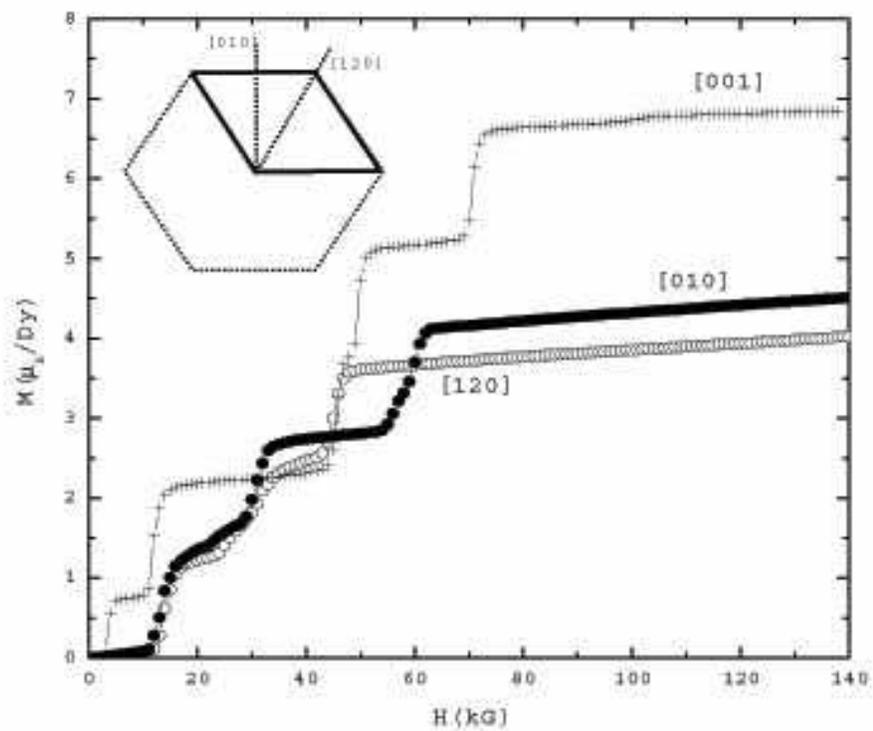}
\end{center}
\caption{Anisotropic magnetization curves shown for three
orientations of the applied field.}\label{f18}
\end{figure}

If we release the restriction that the three Ising systems be
planar, while still imposing that their in-plane projections be
$120^\circ$ apart to agree with the symmetry of the crystals, we
find this modified model can describe the case of DyAgGe quite
well. From the crystal structure of this compound, and the
orthorhombic symmetry of the rare earth sites, we can assume that
the three R magnetic moments of the unit cell align themselves
along three non-planar equivalent directions. We notice that for
$H$=140kG (Fig. \ref{f18}), the ratio of the magnetizations for
the two in-plane orientations (when $H$ is parallel to the two
independent high symmetry axes) is:

\begin{eqnarray}
\frac{M(H\|[120])}{M(H\|[010])}\bigg|_{exp} = \frac{4.02}{4.51} =
0.89 \nonumber
\end{eqnarray}

This is close to the previously calculated value of  for the
planar model assumed for TmAgGe. In the case of DyAgGe, given that
the three $M(H)$ curves in Fig. \ref{f18} seem to indicate that
the moments are tilted outside the $ab$-plane, a model similar to
that used for TmAgGe, enhanced to 3 dimensions, may be
appropriate. Let us assume that the three Dy$^{3+}$ magnetic
moments lie on a cone around the $c$-axis, with the projections in
the $ab$ plane $120^\circ$ apart from each other.  These in-plane
projections will then behave similarly to the full moments in the
case of Tm$^{3+}$, and thus $M(H\|[010])$ represents $\frac{2}{3}$
of the total $ab$ magnetization $M_{ab}$ (if we assume saturation
at $H$=140kG): $M(H\|[010])= \frac{2}{3}M_{ab}$. Then
$M_{ab}=\frac{3}{2} M(H\|[010])= \frac{3}{2}*4.51\mu_B=6.77\mu_B$.
Together with the axial component $M_c(\equiv
M(H\|[001]))=6.85\mu_B$, we can now estimate the total
magnetization at $H$=140kG: $M(140 kG) = \sqrt{(6.77\mu_B)^2 +
(6.85\mu_B)^2} \approx 9.63 \mu_B$, close enough to the calculated
value $\mu_{sat}($Dy$^{3+})=10\mu_B$. We conclude that the model
we assumed for the magnetic moments is consistent with the
experiment. Moreover, one can estimate the angle $\varphi$ of the
saturated magnetization with respect to the $c$-axis: from the
in-plane magnetization being along the $[010]$ direction, we can
conclude that, in the saturated state, the total magnetization
vector lies along a $[ 0 k l ]$ axis. The angle $\varphi$ with the
$c$-axis can be calculated as: $\tan \varphi = \frac{M_{ab}}{M_c}
= \frac{6.77}{6.85} = 0.99$ and thus $\varphi \approx 45^\circ$.

\begin{figure}
\begin{center}
\includegraphics[angle=0,width=120mm]{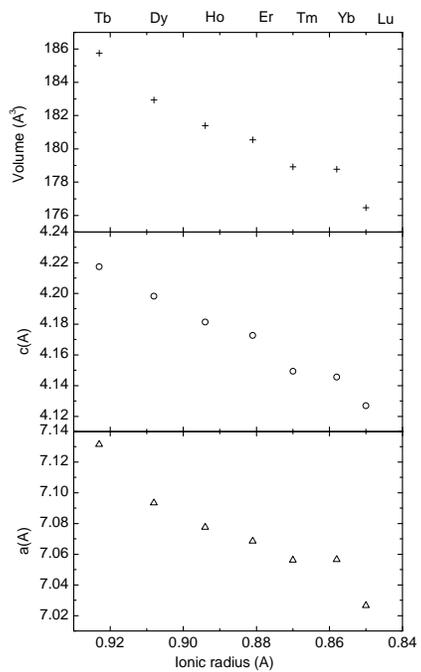}
\end{center}
\caption{Unit cell volumes and lattice parameters for RAgGe,
R=Tb-Lu as a function of R$^{3+}$ ionic radius (bottom axis) or
rare earth (upper axis).}\label{f4}
\end{figure}

The lattice parameters of the hexagonal unit cell in DyAgGe are
given in Fig. \ref{f4} to be $a = 7.09 \AA$ and $c = 4.20 \AA$; in
the orthogonal system of coordinates that we use in defining the
crystalline directions, these correspond to $a' = a =7.09 \AA$,
$b' = a*\frac{\sqrt{3}}{2}= 6.14 \AA$ and $c' = c = 4.20 \AA$. We
can now write the direction given by the angle $\varphi$ in terms
of Miller indices $h$, $k$ and $l$, where $h=0$, and $k$ and $l$
are such that

\begin{eqnarray}
\tan \varphi = 0.99 =\frac{kb'}{lc'} = 1.46 \frac{k}{l}
\Rightarrow \frac{k}{l}=0.68 \nonumber
\end{eqnarray}

The closest integer values for $k$ and $l$ are thus 2 and 3
respectively, which means that in DyAgGe the saturated
magnetization vector $M$ is parallel to the $[ 0 2 3 ]$ direction.

\begin{figure}
\begin{center}
\includegraphics[angle=0,width=120mm]{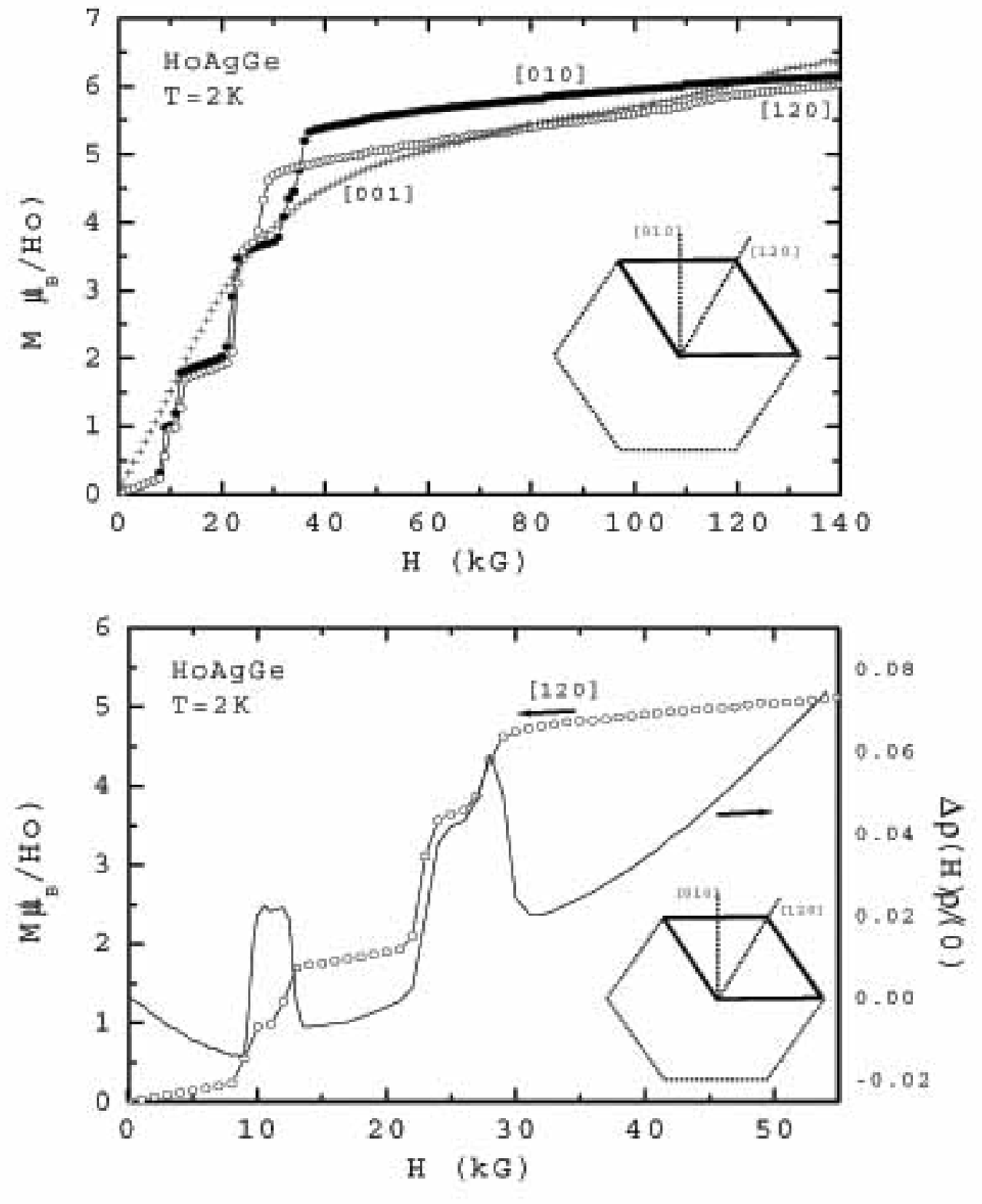}
\end{center}
\caption{(a) Anisotropic magnetization curves in HoAgGe, shown for
three orientations of the applied field at $T$=2K; (b) Transverse
magnetoresistance  and $M(H)$ for $H\|[120]$ at
$T$=2K.}\label{f24}
\end{figure}

HoAgGe resembles DyAgGe in that both these compounds show
metamagnetic transitions for both high-symmetry directions in the
$ab$-plane. However attempts to apply the Dy model to Ho failed
because the in-plane magnetization is almost isotropic in HoAgGe
(Fig. \ref{f24}a), with

\begin{eqnarray}
\frac{M(H\|[120])}{M(H\|[010])}\bigg|_{exp} = \frac{6.01}{6.14} =
0.98 \gg 0.89 = \frac{\sqrt{3}}{2} \nonumber
\end{eqnarray}

Another important observation is that in DyAgGe, the last seen
metamagnetic transition is followed by an almost constant
magnetization (wide plateaus in all three orientations of the
field), supportive of the idea of a stable saturated state above
$\sim$80kG; in HoAgGe however, magnetization keeps increasing with
the applied filed, possibly indicating a continuous spin-flop
transition, as already mentioned. If this is true, than with
fields around 100kG this compound is already in a regime of weak
CEF, therefore any variant of the previous model is inappropriate.
It is worth noting though that HoAgGe is rather curious given its
anisotropy (clear metamagnetism in the two in-plane directions and
no clear transition for $H\|c$). At some future date it would be
interesting to study the angular dependence of metamagnetism in
DyAgGe and HoAgGe, to see how the phase diagrams vary as the
anisotropy is being relaxed.

Insufficient data on ErAgGe (i.e. two in-plane orientations,
higher fields) as well as the low $T_N$ and the single, rather
broad feature in $M(H)$ prevent us from checking how the model
applies to this compound.

All RAgGe described here seem to be good metals, as shown by the
monotonic increase of the resistivity at high temperatures; the
residual resistivity ratio RRR ranges from 2.0 to 6.5, and it did
not seem to improve considerably by annealing in the case of
YbAgGe. However it should be noted that YbAgGe has a RRR of about
10 in a 140kG applied field, significantly larger than that
without field, indicating that a lot of the scattering in this
compound is magnetic in origin.

All of the magnetically ordered compounds in this study, except
for YbAgGe, have a rather unusual, pronounced minimum in the
temperature dependent resistivity above the magnetic transition.
It is similar to the $\rho(T)$ behavior recently found in a
different family, RCuAs$_2$ (for R=Sm, Gd, Tb, Dy) \cite{30}. It
has to be seen if this type of behavior indeed requires, as
suggested in \cite{30}, novel ideas for electrical transport
phenomena in the paramagnetic state of relatively simple magnetic
metals, but the mere fact that similar, atypical, temperature
dependence is observed in resistivity of the members of two
unrelated families, RCuAs$_2$ and RAgGe, in the latter case in
single crystals, certainly asks for some theoretical input as well
as search for other examples.

YbAgGe appears to be a very promising example of a Yb-based
intermetallic compound with clear hybridization.  Whereas it does
not order magnetically within our $M(T)$ measurement range (above
1.85K), it does have a clear small moment ordering below 1.0 K and
also has an enhanced electronic specific heat coefficient value
($\gamma >$ 150 mJ/mol*K$^2$).  Based on these data we can
conclude that YbAgGe is a heavy fermion, with small moment
ordering at very low temperature. Consequently, there is a
possibility of approaching a quantum critical point by experiments
under applied magnetic field, variable pressure, or with changing
lattice parameters with doping, or with variable concentration of
the magnetic moments (i.e. of the Yb$^{3+}$ ions).  While in an
isotropic case, application of pressure in Yb compounds close to
quantum critical point is expected to increase the ordering
temperature \cite{31}, crystallographically and electronically
anisotropic materials like YbAgGe may have non-trivial response to
pressure. More investigation is needed for a better understanding
of the nature of the ordering in YbAgGe.

{\it Acknowledgments:} We would like to thank Cedomir Petrovic and
Raquel Ribeiro for help with powder X-ray diffraction and Cathie
Condron for single crystal X-ray analysis of Gd$_3$Ag$_4$Ge$_4$.
Ames Laboratory is operated for the U.S. Department of Energy by
Iowa State University under Contract No. W-7405-Eng.-82. This work
was supported by the Director for Energy Research, Office of Basic
Energy Sciences. Work at NHMFL - Los Alamos Facility was performed
under auspices of the National Science Foundation and the U.S.
Department of Energy.

\clearpage

\clearpage

\clearpage

\clearpage

\clearpage

\clearpage

\clearpage

\clearpage

\clearpage

\clearpage

\clearpage

\clearpage

\clearpage

\clearpage

\clearpage

\clearpage

\clearpage

\clearpage

\clearpage

\clearpage

\clearpage

\clearpage

\clearpage

\clearpage

\clearpage

\clearpage

\clearpage

\clearpage

\clearpage

\clearpage

\clearpage

\clearpage

\clearpage

\clearpage

\clearpage

\clearpage

\clearpage

\clearpage

\clearpage

\clearpage

\clearpage

\end{document}